\documentclass[12pt]{article}

\usepackage{times}
\usepackage[super,comma]{natbib}
\usepackage{mathtools}
\usepackage{amssymb}
\usepackage{amsmath}
\usepackage{amsfonts}
\usepackage{graphicx}
\usepackage{xcolor}

\usepackage[colorlinks,
            citecolor=blue,
            linkcolor=red,
            urlcolor=blue]{hyperref}

\title{Detection of Kardar-Parisi-Zhang hydrodynamics in a quantum Heisenberg spin-$1/2$ chain}

\author{
A. Scheie,$^{1}$ N.E. Sherman,$^{2,3}$ M. Dupont,$^{2,3}$\\
S.E. Nagler,$^{1}$ M.B. Stone,$^{1}$ G.E. Granroth,$^{1}$ J.E. Moore,$^{*2,3}$ D.A. Tennant$^{*4,5,6}$\\
\\
\normalsize{$^{1}$Neutron Scattering Division, Oak Ridge National Laboratory,}\\
\normalsize{Oak Ridge, Tennessee 37831, USA}\\
\normalsize{$^{2}$Department of Physics, University of California,}\\
\normalsize{Berkeley, California 94720, USA}\\
\normalsize{$^{3}$Materials Sciences Division, Lawrence Berkeley National Laboratory,}\\
\normalsize{Berkeley, California 94720, USA}\\
\normalsize{$^{4}$Materials Science and Technology Division, Oak Ridge National Laboratory,}\\
\normalsize{Oak Ridge, Tennessee 37831, USA}\\
\normalsize{$^{5}$Shull-Wollan Center, Oak Ridge National Laboratory,}\\
\normalsize{Oak Ridge, Tennessee 37831, USA}\\
\normalsize{$^6$Quantum Science Center,}\\
\normalsize{Oak Ridge, Tennessee 37831, USA}\\
\\
\normalsize{$^\ast$To whom correspondence should be addressed:}\\
\normalsize{jemoore@berkeley.edu, tennantda@ornl.gov}
}

\date{}

\begin{document}

\setcounter{page}{0}
Notice: This manuscript has been authored by UT-Batelle, LLC, under contract DE-AC05-00OR22725 with the US Department of Energy (DOE). The US government retains and the publisher, by accepting the article for publication, acknowledges that the US government retains a nonexclusive, paid-up, irrevocable, worldwide license to publish or reproduce the published form of this manuscript, or allow others to do so, for US government purposes. DOE will provide public access to these results of federally sponsored research in accordance with the DOE Public Access Plan (http://energy.gov/downloads/doe-public-access-plan).

\newpage

\baselineskip 24pt 

\maketitle

\newpage
\begin{quote}\bfseries
    Classical hydrodynamics is a remarkably versatile description of the coarse-grained behavior of many-particle systems once local equilibrium has been established~\cite{landau1987}. The form of the hydrodynamical equations is determined primarily by the conserved quantities present in a system. Some quantum spin chains are known to possess, even in the simplest cases, a greatly expanded set of conservation laws, and recent work suggests that these laws strongly modify collective spin dynamics even at high temperature~\cite{castroalvaredo2016,bertini2016}. Here, by probing the dynamical exponent of the one-dimensional Heisenberg antiferromagnet KCuF$_3$ with neutron scattering, we find evidence that the spin dynamics are well described by the dynamical exponent $z=3/2$, which is consistent with the recent theoretical conjecture that the dynamics of this quantum system are described by the Kardar-Parisi-Zhang universality class~\cite{kardar1986,ljubotina2019}. This observation shows that low-energy inelastic neutron scattering at moderate temperatures can reveal the details of emergent quantum fluid properties like those arising in non-Fermi liquids in higher dimensions.
\end{quote}

Interacting magnetic moments (``spins'') governed by the laws of quantum mechanics can exhibit a vast set of complex phenomena such as Bose-Einstein condensation and superfluidity~\cite{giamarchi2008}, topological states of matter~\cite{haldane1983}, and exotic phase transitions~\cite{sachdev2008,faure2018}. Understanding quantum magnets is therefore a challenging task, connecting experiments with intensive theoretical modeling and state-of-the-art numerical simulations. In that respect, linear arrangements of spins (``spin chains'') at temperatures close to absolute zero have been influential because the one-dimensional ($1$D) setting produces especially prominent quantum fluctuations~\cite{giamarchi2004}.

The most celebrated model magnetic system realized in nature is the Heisenberg spin-half chain, where isotropic magnetic moments are coupled by a nearest-neighbor antiferromagnetic exchange interaction of strength $J$. It is characterized by fractional quasiparticles excitations called spinons (Fig.~\ref{flo:Schematic}b) with a dispersion relation given by $\hbar\omega(Q)=J\frac{\pi}{2}|\sin(Qa)|$, where $a$ is the lattice spacing unit (following convention for $1$D chains, we set $a=1$ in this study). They are responsible for the physical properties of the system and can be identified by the dynamical spin response function, as measured in inelastic neutron spectroscopy. Spinons are created in pairs, leading to a continuum in the neutron scattering spectrum, and interact with one another. In fact, in the ground state, two-spinon states accounting for $71\%$ of the total spectral weight have an upper bound $\hbar\omega(Q)=J\pi|\sin(Q/2)|$, which gives its distinctive shape to the spectrum (Fig.~\ref{flo:ExperimentRange}a), and including the four-spinon contribution on top of the two-spinon exhausts $98\%$ of the weight~\cite{Caux_2006}, etc.

As temperature increases and many spins are excited, the spin dynamics at frequencies $\hbar\omega\ll k_\mathrm{B}T$ is usually thought of in terms of collective thermal rather than quantum effects. This high-temperature regime has not been the focus of experimental study, but recent theoretical progress in $1$D quantum systems suggests that it nevertheless holds precious information on the underlying quantum features~\cite{castroalvaredo2016,bertini2016,Bulchandani2018}. One can make an analogy with the phenomenological derivation of the equations of fluid dynamics, based on the continuity equations of conserved quantities (such as mass, energy, or momentum): depending on the intrinsic quantum conservation laws of the system, one expects the emergence of different kinds of coarse-grained hydrodynamic behaviors for the spins at high-temperature. Remarkably, some $1$D quantum systems, known as integrable --- including the Heisenberg spin-half chain --- possess an infinite number of nontrivial conserved quantities. They strongly constrain the overall dynamics of integrable systems and can endow them with peculiar hydrodynamic properties, some of which have been observed experimentally in a $1$D cloud of trapped~$^{87}$Rb atoms~\cite{schemmer2019}. In the case of magnets, three universal regimes have been identified~\cite{dupont2020,denardis2020} and are classified by the dynamical exponent $z$, governing the length-time scaling, i.e., $\textsc{length}\sim\textsc{time}^{1/z}$: $z=2$ corresponds to diffusion, $z=1$ to ballistic, and $z=3/2$ to superdiffusive dynamics (Fig.~\ref{flo:Schematic}c).

The presence of ballistic spin dynamics in integrable systems is theoretically established by showing that at least part of the spin current $\hat{j}_\mathrm{s}$ in an initial state persists to infinite time, resulting in an infinite spin dc conductivity. Quantitatively, this can be achieved by looking at the long-time asymptote of the spin current-current correlation $\lim_{t\to\infty}\left\langle\hat{j}_\mathrm{s}(t)\hat{j}_\mathrm{s}(0)\right\rangle$, where saturation to a nonzero value signals ballistic spin transport and a nonzero Drude weight. Although challenging for many-body quantum systems, the Drude weight can be accessed numerically~\cite{karrasch2012} and a lower bound can often be obtained analytically~\cite{zotos1997,prosen2011}.  Diffusive behavior, one of the other universal regimes, is typically recovered for systems with zero Drude weight, which implies eventual relaxation of spin currents and finite transport coefficients. Unexpectedly, an intermediate scenario was recently unveiled~\cite{ilievski2018,ljubotina2019,gopalakrishnan2019,denardis2019}: a zero Drude weight but a slowly decaying (typically algebraically with time) spin current-current correlation, giving rise to superdiffusive dynamics with $z=3/2$. The intermediate scenario was found numerically~\cite{ljubotina2019} by calculation of the full scaling function to belong to the Kardar-Parisi-Zhang (KPZ) universality class in 1+1 dimension, reproduced in the Supplementary Material; a theoretical scenario for how KPZ dynamics emerges in the Heisenberg chain has been proposed~\cite{Bulchandani2020}.

This universality class originates from the classical non-linear stochastic partial differential equation of the same name~\cite{kardar1986}, initially introduced to describe the evolution in time of the profile of a growing interface. Generally speaking, a system is considered to be in the KPZ universality class if its long-time dynamics shows the same scaling laws as appear in the KPZ equation itself. Besides interface growth~\cite{takeuchi2010}, such scaling has been found in disordered conductors~\cite{somoza2007}, quantum fluids~\cite{kulkarni2015}, quantum circuits~\cite{nahum2017}, traffic flow~\cite{degier2019}, and was recently predicted  also to appear in the high-temperature dynamics of some one-dimensional integrable quantum magnets~\cite{ljubotina2019,denardis2019,dupont2020,denardis2020}, although its exact microscopic origin is still under active research in this case.

Here, using neutron scattering experiments on KCuF$_3$, which realizes a nearly ideal quantum Heisenberg spin-half chain, we report on the observation of KPZ dynamics at various temperatures in the range $T=75$~K to $T=300$~K. Combining experimental measurements with extensive numerical simulations based on a microscopic description of the system, we identify a characteristic power-law $\propto Q^{-3/2}$ behavior in the neutron scattering spectrum, in agreement with the KPZ universality class predictions~\cite{prahofer2004}.

\section*{Searching for Kardar-Parisi-Zhang hydrodynamics}

\begin{figure}
	\centering
	\includegraphics[width=\textwidth]{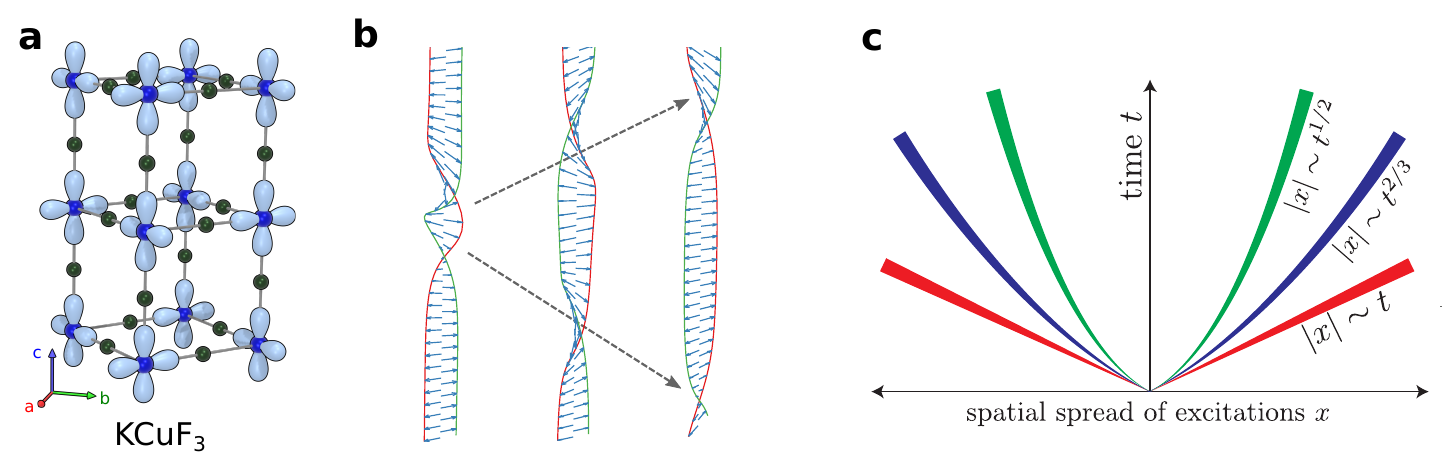}
	\caption{\textbf{\textsf{a}} Crystal structure of KCuF$_3$, showing the orbital order of the Cu $x^2-y^2$ orbitals. This order leads to strong magnetic exchange interactions along the $c$ (vertical) axis and weak exchange interactions along $a$ and $b$, such that the Cu$^{2+}$ ions effectively make 1D chains. \textbf{\textsf{b}} Schematic illustration of spinon excitations in a 1D Heisenberg antiferromagnet (based on Ref.~\cite{haldane1983}). \textbf{\textsf{c}} Schematic illustration of three possible length-time scaling behaviors $|x|\sim t^{1/z}$ observed at high temperature in $1$D quantum magnets, classified by the dynamical exponent $z$: $z=2$ corresponds to diffusion (green curve), $z=3/2$ to superdiffusive (blue curve) and $z=1$ to ballistic dynamics.}
	\label{flo:Schematic}
\end{figure}

KCuF$_3$ has long been studied as a model of $1$D Heisenberg antiferromagnetism with $S=1/2$ spins borne by Cu$^{2+}$ ions~\cite{Hirakawa1970,Tennant1993,Lake2013}. Due to the Cu$^{2+}$ $d_{x^2-y^2}$ orbital order (Fig.~\ref{flo:Schematic}a), the magnetic exchange interaction is limited to nearest-neighbor spins and is spatially anisotropic. It is dominant along the $c$ axis ($J_c=33.5$~meV) while the interchain coupling is much weaker ($J_{a,b}=-1.6$~meV), leading to effective one-dimensional $c$ axis spin-half chains. Although the system magnetically orders at $T_N=39$~K due to the inherent presence of a finite exchange interaction $J_{a,b}$, its behavior for $T\gtrsim T_N$ is a good approximation to an ideal $1$D Heisenberg antiferromagnet which can be modeled by the following Hamiltonian,
\begin{equation}
    \hat{\mathcal{H}}=J_c\sum\nolimits_{n}\hat{\boldsymbol{S}}_{n}\cdot\hat{\boldsymbol{S}}_{n+1},
    \label{eq:hamiltonian}
\end{equation}
with $\hat{\boldsymbol{S}}_{n}$ the spin-$1/2$ operator on the site index $n$. At equilibrium, the spin dynamics can be characterized through the correlation function between two spatially separated spins at different moments in time, and whose Fourier transform to momentum and frequency spaces is the dynamical spin structure factor $\mathcal{S}\bigl(Q,\omega\bigr)$. This quantity is directly proportional to the measured inelastic neutron scattering intensity, and can be computed numerically for the model~\eqref{eq:hamiltonian} using matrix product state (MPS) techniques (See Methods), allowing for a direct comparison between theory and experiments. Especially, the universal dynamical exponent $z$ is expected to manifest itself in the form~\cite{prahofer2004},
\begin{equation}
    \mathcal{S}\bigl(Q,\omega{\to 0}\bigr)\sim Q^{-z},
    \label{eq:hydro_scaling}
\end{equation}
in the limit of small momentum $Q$ and vanishing energy $\hbar\omega$, with KPZ behavior identified by $z=3/2$.

\begin{figure}
	\centering
	\includegraphics[width=0.9\textwidth]{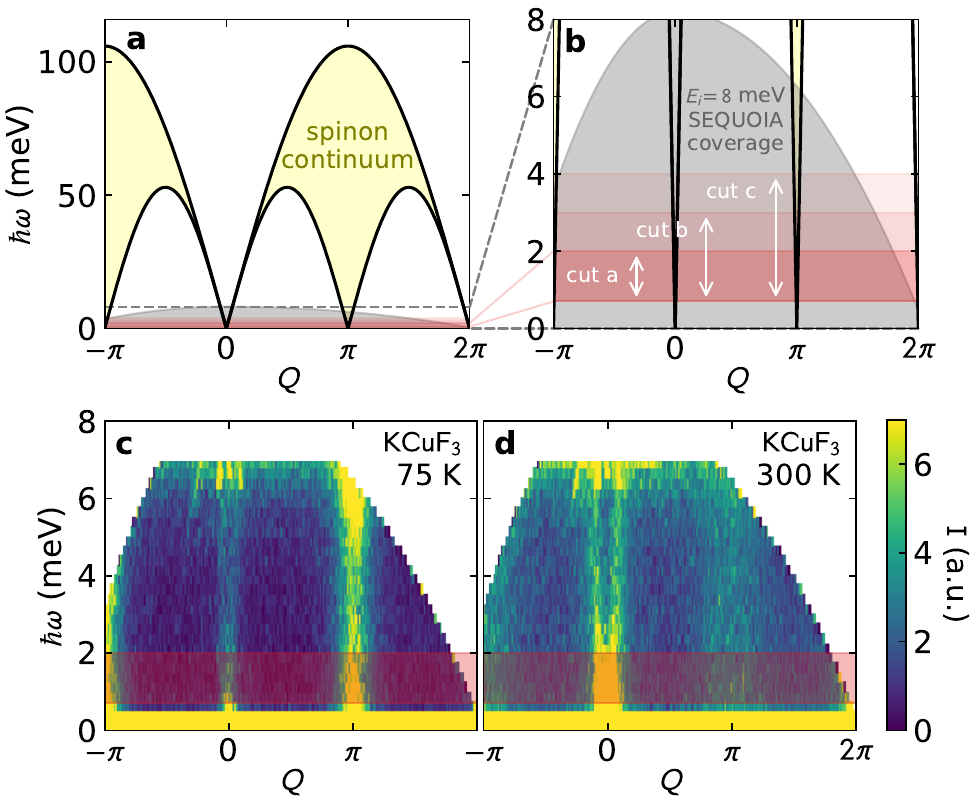}
	\caption{Measured neutron spectrum of KCuF$_3$. \textbf{\textsf{a}} Cartoon of the KCuF$_3$ spinon spectrum. The gray region at the bottom shows the region measured. \textbf{\textsf{b}} Zoom in on the region measured in the SEQUOIA experiment, also showing three cuts (cut a, cut b, and cut c) used to approximate the $\hbar\omega\to0$ scattering. \textbf{\textsf{c}} and \textbf{\textsf{d}} show measured spectra at $75$~K and $100$~K, respectively. Cut a is indicated by the horizontal red bar. It is not possible to directly measure the magnetic scattering at $\hbar\omega\to0$ due to the strong elastic incoherent scattering. Therefore, we take the lowest energy cuts where magnetic scattering dominates, cut a, as shown in Fig. \ref{flo:PLfits}.}
	\label{flo:ExperimentRange}
\end{figure}

To search for effects of KPZ behavior, we measured the inelastic KCuF$_3$ neutron spectrum using the SEQUOIA spectrometer at the Spallation Neutron Source at Oak Ridge National Laboratory. Our sample was a $6.86$~g KCuF$_3$ single crystal mounted with the $c$ axis perpendicular to the incident beam. To probe hydrodynamic signatures, we focused on the low-energy part of the spectrum. We measured with an incident energy $E_i=8$~meV, which gives access to the very bottom of the spectrum (the total KCuF$_3$ bandwidth is $105$~meV~\cite{Lake2013}), as shown in Fig.~\ref{flo:ExperimentRange}, with a resolution full width at half maximum $0.25$~meV. Elastic incoherent scattering prevents us from isolating the magnetic scattering at $\hbar\omega\to0$, so we take the $0.7<\hbar\omega<2$~meV scattering to be an approximation to the $\hbar\omega\to0$ spectrum. To evaluate the robustness of this approximation, we consider three different energy ranges with $\hbar\omega>0.7$~meV (this is empirically where elastic incoherent scattering background is negligible), as indicated in Fig.~\ref{flo:ExperimentRange}. They all lead to similar results (see Supplementary Information for details).

\section*{Results}

\begin{figure}
	\centering
	\includegraphics[width=0.9\textwidth]{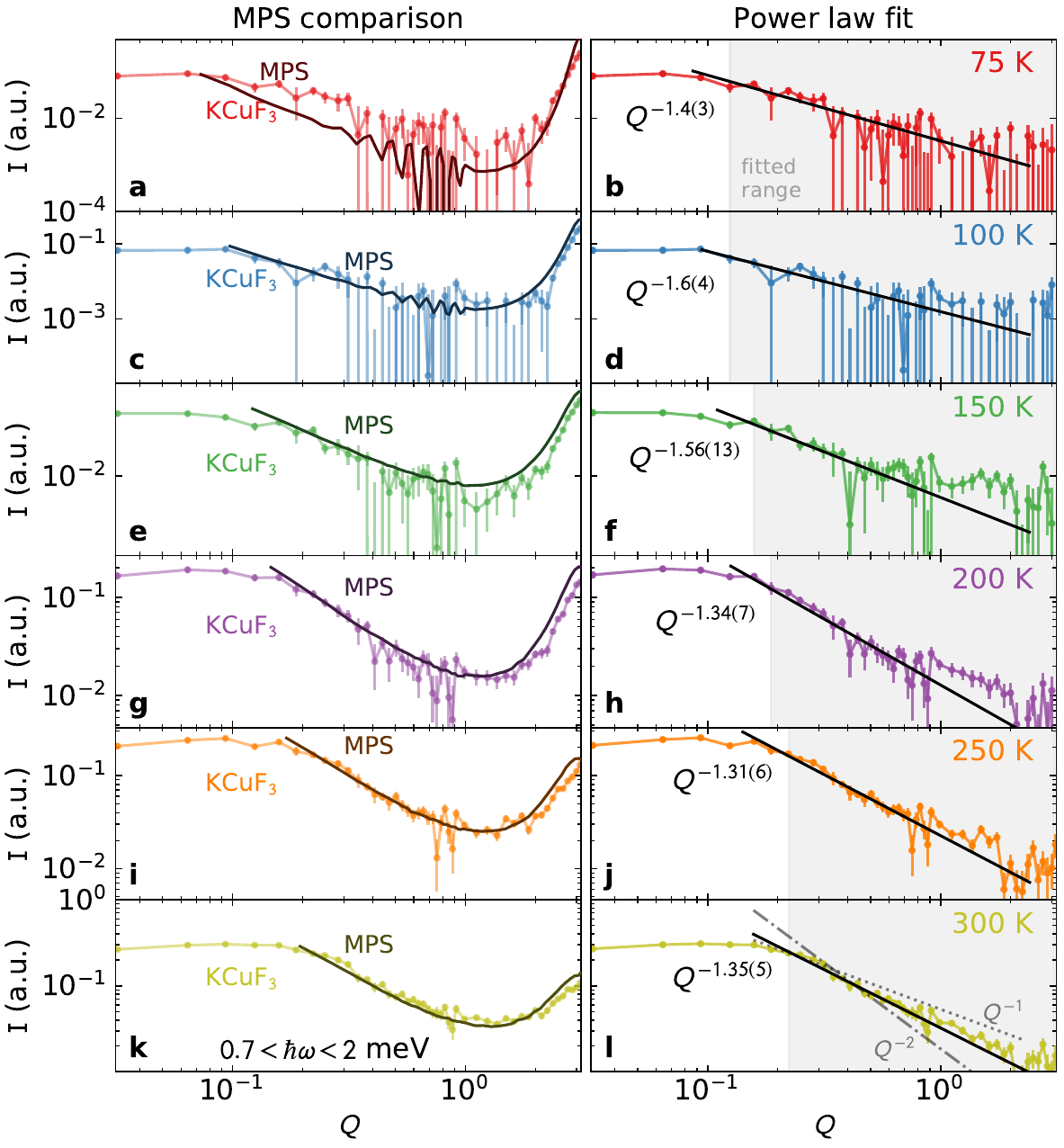}
	\caption{Power law behavior of KCuF$_3$ around $Q=0$. The left column shows experimental data integrated over $0.7<\hbar\omega<2$ meV (cut a in Fig.~\ref{flo:ExperimentRange}) symmetrized about $Q=0$ compared with the MPS simulations. The same multiplicative scaling factor is used for all temperatures, and the agreement is quite good above $Q\approx0.2$, below which finite-size effects are significant for MPS (See the Supplementary Information). The right column shows the data fitted to a phenomenological power law. As a part of the fit, the $Q=\pi$ peak was also fitted to a power law and subtracted off as background. The fitted power is very close to $-3/2$ at all temperatures. Comparison to $z=2$ and $z=1$ exponents are given in panel \textbf{\textsf{l}}.  (Note that $Q$ is unitless $0 \to 2\pi$ as in Fig. \ref{flo:ExperimentRange}.)}
	\label{flo:PLfits}
\end{figure}

A quantitative test to distinguish different kinds of hydrodynamics is the scattering intensity behavior at small energy versus $Q$ near the ferromagnetic wavevector $Q=0$, see Eq.~\eqref{eq:hydro_scaling}. We first compare these data to MPS simulations integrated over the same energy range in Fig.~\ref{flo:PLfits}. There is a very good agreement between the two, with deviations only appearing at the low temperature (mainly $75$~K). This discrepancy is attributed to the inherent interchain couplings (not present in the pure $1$D model), and which lead to antiferromagnetic ordering at $T_N=39$~K. In fact, it was theoretically shown~\cite{dupont2018} that for dynamical quantities, the $1$D temperature crossover in quasi-one-dimensional systems such as KCuF$_3$ is spoiled for $T\lesssim 3T_N$. Furthermore the spectral intensity at low-energy in a strictly $1$D system away from $Q\to\pm 0$ and $\pm\pi$ is greatly suppressed with temperature, making an accurate estimation from numerical simulations difficult; hence the non-physical oscillatory behavior in the MPS data of Fig.~\ref{flo:PLfits} at $100$~K and $75$~K for intermediate $Q$ values. The $Q\lesssim 0.1$ experimental data deviates from power law behavior, partly because of $Q$ resolution broadening, and partly because of the dispersion peaking away from $Q=0$ at finite energy. Moreover, the numerical simulations do not allow us to reliably access the $Q\lesssim 0.2$ regime: this is because simulations are performed on finite-length chains (typically a hundred spins on the lattice) which introduces an artificial cutoff at low $Q$ as $\hbar\omega\to 0$ when it comes to the dynamics, as compared to a system in the thermodynamic limit (See the Supplementary Information). 

Having identified the temperature window where $1$D physics take place, we consider in Fig.~\ref{flo:PLfits} the same experimental data from which a phenomenological $Q=\pi$ power-law background is subtracted. This highlights the $Q\to0$ regime, where power-law fits of the form $\propto Q^{-z}$ give an exponent close to $z\approx1.5$ at all temperatures. Note that fits without this phenomenological background yield results which are equivalent to within uncertainty, and are shown in the Supplementary Information. (Also note that experimental resolution broadening increases the fitted power by 2-3\%, shown in the Supplemental Information, so the true 300~K exponent may be closer to $Q^{-1.31(5)}$.) This correction notwithstanding, the fitted exponent, while not $Q^{-3/2}$ exactly, is clearly closer to that value than to either $Q^{-1}$ (ballistic) or $Q^{-2}$ (diffusive). See Fig.~\ref{flo:PLfits}l for a comparison between the three cases. At $T=300$~K, the discrepancy between the expected $Q^{-3/2}$ behavior and the experimental fit $Q^{-1.35(5)}$ can be explained by the fact that we are not measuring at exactly $\hbar\omega=0$. 
Indeed, we show in the Supplementary Information that finite-energy fits to the MPS data also yield powers less than $z=3/2$, while fits in the low-energy limit $\hbar\omega\to0$ --- more amenable numerically than experimentally --- precisely deliver the correct exponent. This effect is experimentally unavoidable, and we thus consider our experimental results, in conjunction with a comprehensive numerical study observing the same trend, to be clearly more consistent with the KPZ universality class behavior than either of the conventional possibilities, ballistic or diffusive behavior.  A different experiment (e.g., with polarized neutrons or a spin-echo spectrometer) may be able to measure the magnetic scattering at the elastic line.

\begin{figure}
	\centering
	\includegraphics[width=0.99\textwidth]{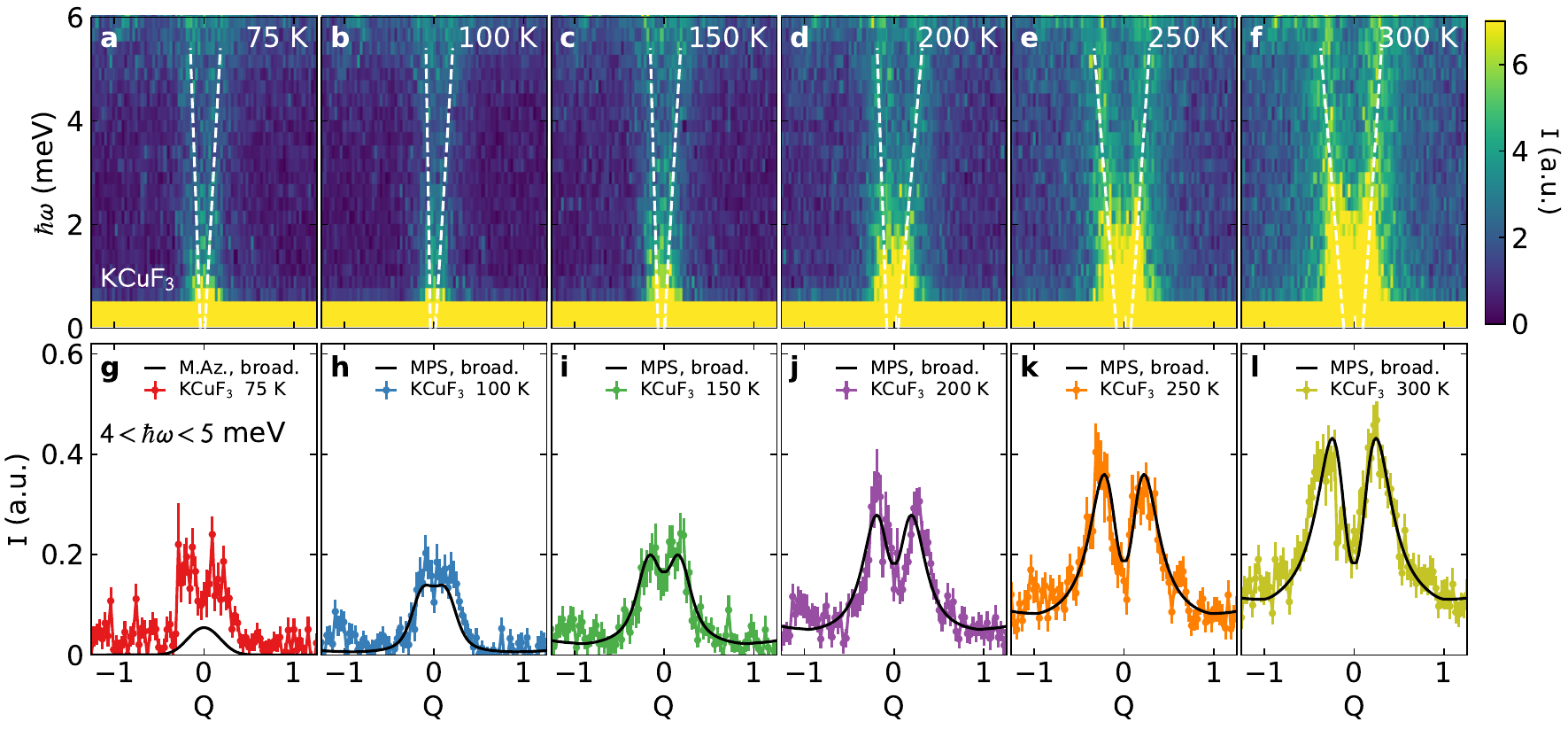}
	\caption{Temperature evolution of the KCuF$_3$ neutron spectra around $Q=0$. ($Q$ units are defined in Fig. \ref{flo:ExperimentRange}.) Panels \textbf{\textsf{a}} - \textbf{\textsf{f}} show colormap plots of the spectra (white dashed lines are fitted linear dispersions), and panels \textbf{\textsf{g}} - \textbf{\textsf{l}} show constant energy cuts of $4<\hbar\omega<5$~meV. The lower panel also includes theoretical curves for comparison. (Panel \textbf{\textsf{g}} shows the zero-temperature M\"uller ansatz~\cite{muller1981} scaled to match the $Q=\pi$ intensities, the rest show the MPS calculations. Resolution broadening has been applied to the theoretical curves, see Supplementary Information for details.) These show the spinon modes splitting more as temperature increases, indicating a significant mode softening.}
	\label{flo:Spectra}
\end{figure}

In measuring the low-energy KCuF$_3$ neutron spectrum, we also noticed a previously unreported feature, shown in Fig.~\ref{flo:Spectra}: as temperature increases, the dispersion around $Q=0$ softens. In other words, the split between $\pm Q$ modes increases, showing a decreased dispersion velocity: $190(20)$~meV$\cdot$\AA~at $100$~K (within uncertainty of the theoretical $T=0$ velocity $207$~meV$\cdot$\AA) to $84(9)$~meV$\cdot$\AA~at $300$~K. This feature is also captured by the MPS simulations, as shown in Fig.~\ref{flo:Spectra}g-l. This temperature dependent mode softening has not been noticed before, and indicates that the excitation velocity is renormalized by interaction with other quasiparticles; such spin wave velocity renormalization also exists in the two-dimensional quantum Heisenberg antiferromagnet.\cite{Chakravarty1989} We leave it an open question whether this mode softening may relate to KPZ physics or be an apparent shift from broadening by material-dependent damping.

\section*{Conclusion}

Our results experimentally show the presence of KPZ-like physics in KCuF$_3$, characterized by the dynamical exponent $z=3/2$. The observation of this superdiffusive scaling in the high-temperature dynamics of a Heisenberg chain demonstrates that inelastic neutron scattering can complement transport as a probe of quantum coherent collective phenomena, even when those phenomena do not have an interpretation in terms of a small number of quasiparticles. In some situations transport measurements are prohibitively difficult, like in many non-chiral $1$D systems where even trace impurities or small defects simply interrupt macroscopic transport. In such cases, neutron scattering can probe subtle fluid properties where the dynamical exponent $z$ around the low-energy dispersion reveals the nature of the collective quasiparticle flow.

In higher dimensions, this approach can provide insight into analogues of non-Fermi liquid scaling and other behaviors hypothesized to exist from transport, with the advantage of an increased degree of quantum coherence and a precise characterization of the model Hamiltonian. In one dimension, this work goes significantly beyond previous efforts with other techniques to see generalized hydrodynamics from integrability by revealing a different scaling regime. Theoretically, there remain important questions to be understood, such as how to characterize the crossover regime when integrability is broken by interchain couplings or other residual interactions. In particular, there are spin chains where the field-theory description at low temperatures shows emergent integrability but the lattice-scale physics is not integrable, unlike for the Heisenberg model, and it should be possible to apply our combined experimental and computational approach to this category of systems as well.

\section*{Methods}

Full methods are available in the Supplementary Information.

\subsection*{Inelastic neutron scattering}

Neutron scattering measurements were performed using the SEQUOIA spectrometer\cite{Granroth2006,Granroth2010} on a single crystal mounted in the $hh\ell$ scattering plane, such that the $c$ axis is perpendicular to the beam. Although this does not measure a full reciprocal space volume, we integrate over the $h$ and $k$ directions because the magnetic signal only depends upon $\ell$. At $E_i=8$~meV, we measure below the lowest phonon dispersion, so the only contributions to the signal near $Q=0$ are magnetic scattering and elastic background. We corrected the magnetic scattering for the form factor by calculating the magnitude of $Q$ for each pixel and dividing by the Cu$^{2+}$ form factor, as explained in the Supplementary Information. Data were plotted using Mslice~\cite{DAVE}.

\subsection*{Numerical simulations}

The numerical simulations are based on a matrix product state (MPS) approach~\cite{schollwock2011} using the ITensor library~\cite{itensor} to simulate the finite-temperature spin dynamics of the $1$D quantum Heisenberg spin-half chain. We use the purification method to represent the finite-temperature quantum state (mixed state) as a pure state in an enlarged Hilbert space~\cite{verstraete2004}. For the dynamics, we use a method expanding the spectral function $\mathcal{S}(Q,\omega)$ in terms of Chebyshev polynomials~\cite{holzner2011}.

\section*{Data Availability}
 All plotted experimental data are publicly available at \href{doi.org/10.13139/ORNLNCCS/1668822}{doi.org/10.13139/ORNLNCCS/1668822}.

\section*{Acknowledgments}

N.S., M.D. and J.E.M. were supported by the U.S. Department of Energy, Office of Science, Office of Basic Energy Sciences, Materials Sciences and Engineering Division under Contract No. DE-AC02-05-CH11231 through the Scientific Discovery through Advanced Computing (SciDAC) program (KC23DAC Topological and Correlated Matter via Tensor Networks and Quantum Monte Carlo). During the last period of the work, N.S. was supported by the U.S. Department of Energy, Office of Science, Office of Basic Energy Sciences, Materials Sciences and Engineering Division under Contract No. DE-AC02-05-CH11231 through the Theory Institute for Molecular Spectroscopy (TIMES).  J.E.M. was also supported by a Simons Investigatorship. This research used resources at the Spallation Neutron Source, a DOE Office of Science User Facility operated by the Oak Ridge National Laboratory. D.A.T. and J.E.M. were supported by the U.S. Department of Energy, Office of Science, National Quantum Information Science Research Centers.
This research also used the Lawrencium computational cluster resource provided by the IT Division at the Lawrence Berkeley National Laboratory (Supported by the Director, Office of Science, Office of Basic Energy Sciences, of the U.S. Department of Energy under Contract No. DE-AC02-05CH11231). This research used resources of the Compute and Data Environment for Science (CADES) at the Oak Ridge National Laboratory, which is supported by the Office of Science of the U.S. Department of Energy under Contract No. DE-AC05-00OR22725. This research also used resources of the National Energy Research Scientific Computing Center (NERSC), a U.S. Department of Energy Office of Science User Facility operated under Contract No. DE-AC02-05CH11231.

\section*{Author contributions}

A.S. and N.E.S. contributed equally to this work. The project was conceived by J.E.M. and D.A.T.. The experiments were performed by A.S., M.B.S., S.E.N., and D.A.T.. The numerical simulations and theoretical analysis were performed by N.E.S., M.D., and J.E.M.. The neutron data analysis was performed by A.S. with input from D.A.T., G.E.G., and S.E.N.. The results were discussed and paper written by A.S., N.E.S., M.D., S.E.N., J.E.M and D.A.T..

\section*{Additional information}

Correspondence and requests for materials should be addressed to J.E. Moore and D.A. Tennant.

\section*{Competing financial interests}

The authors declare no competing financial interests.

\newpage

\renewcommand{\thefigure}{S\arabic{figure}}
\renewcommand{\thetable}{S\arabic{table}}
\renewcommand{\theequation}{S.\arabic{equation}}
\renewcommand{\thepage}{S\arabic{page}}  
\setcounter{figure}{0}
\setcounter{equation}{0}
\setcounter{page}{1}

\baselineskip18pt 

\baselineskip18pt 

{\centering
	\section*{Supplementary Information for Detection of Kardar-Parisi-Zhang hydrodynamics in a quantum Heisenberg spin-$1/2$ chain}}

\vspace{1in}

{
	\hypersetup{linkcolor=black}
	\tableofcontents
}

\newpage

\section{Experiments}
\subsection{Data collection}

\begin{figure}
	\centering
	\includegraphics[width=\textwidth]{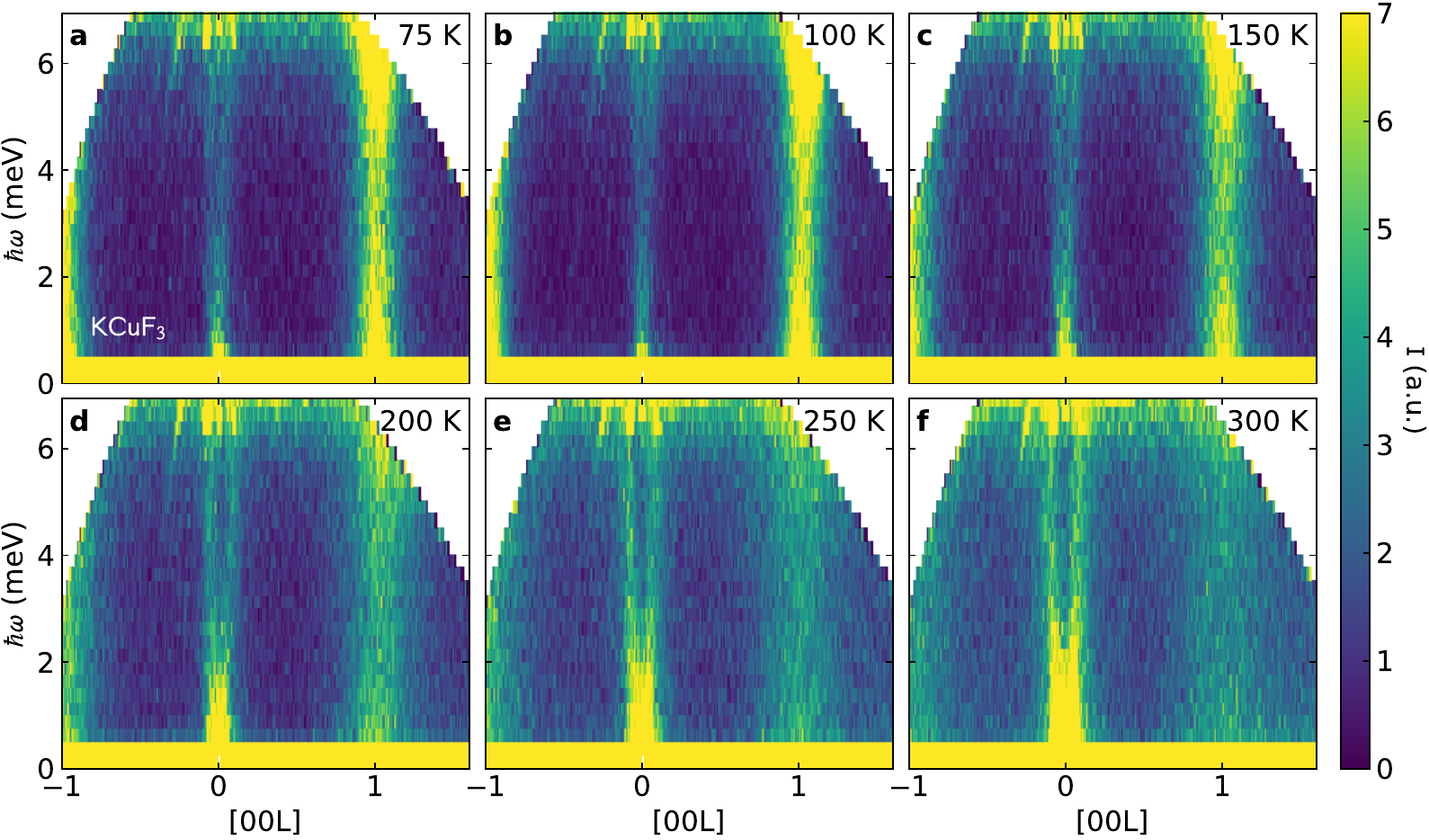}
	\caption{Temperature dependent neutron spectra of KCuF$_3$.}
	\label{flo:FullSpectra}
\end{figure}

\begin{figure}
	\centering
	\includegraphics[width=0.9\textwidth]{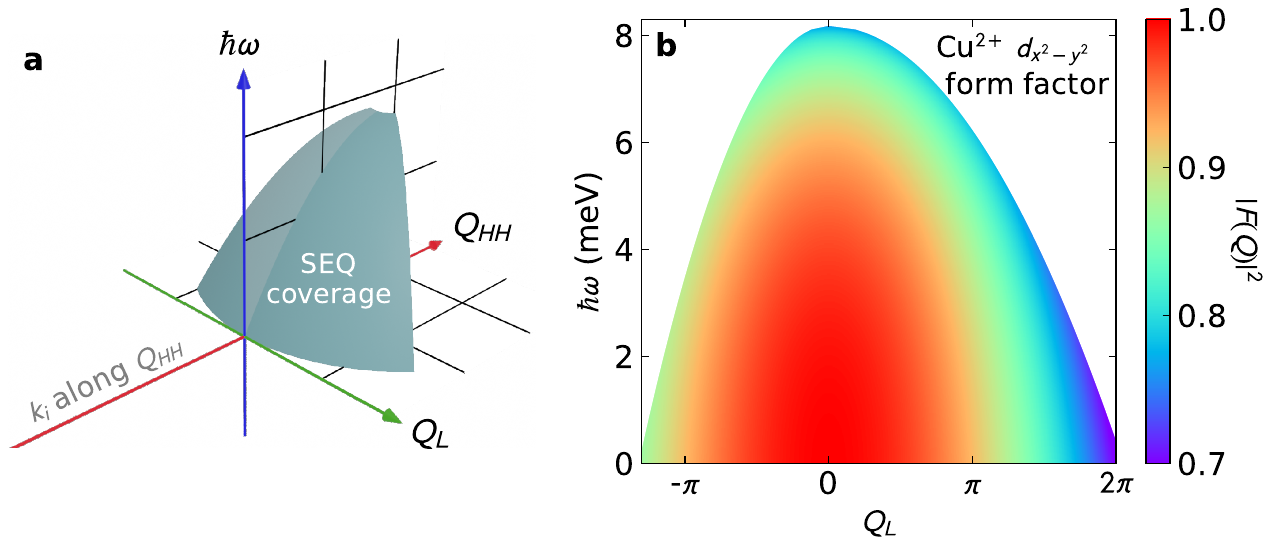}
	\caption{Form factor correction for KCuF$_3$ SEQUOIA experiment. Panel \textbf{\textsf{a}} shows the coverage in reciprocal space and energy. Because of the instrument geometry, any finite energy or momentum transfer also involves a finite momentum transfer along the $(h,h,0)$ direction. This leads to the Cu$^{2+}$ $d_{x^2-y^2}$ form factor correction shown in panel \textbf{\textsf{b}}, which is calculated based off $\Vec{Q}$.}
	\label{flo:FF_correction}
\end{figure}

\begin{figure}
		\centering
		\includegraphics[width=0.85\textwidth]{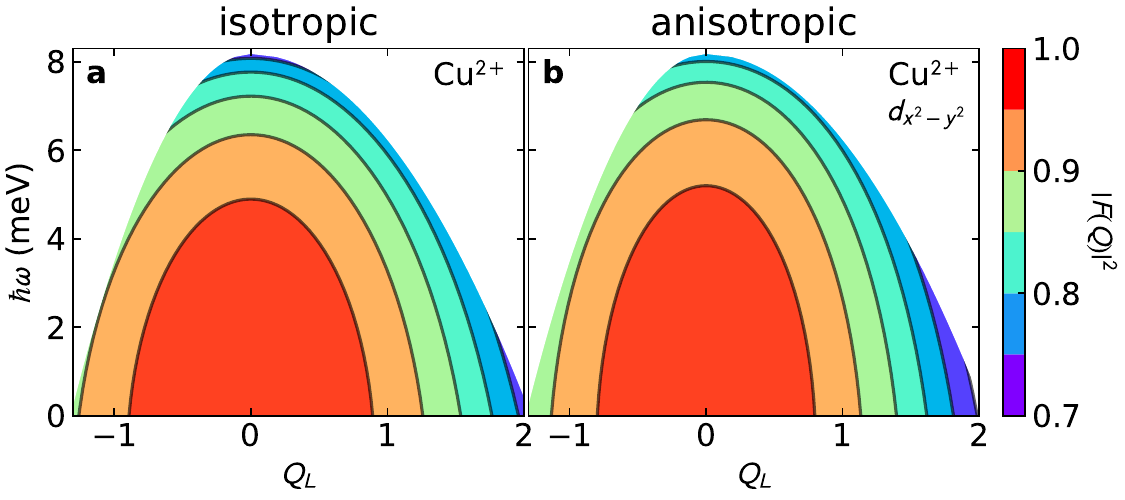}
		\caption{\textbf{\textsf{a}} Color contour map of the isotropic Cu$^{2+}$ form factor compared to \textbf{\textsf{b}} the anisotropic $d_{x^2-y^2}$ Cu$^{2+}$ form factor for KCuF$_3$. The differences are subtle: the anisotropic form factor falls off faster with $Q_{L}$ but more gradually with $\hbar \omega$ than the isotropic form factor.}
		\label{flo:FF_anisotropy}
	\end{figure}

We measured six temperatures of KCuF$_3$ with the $c$ axis perpendicular to the incident neutron beam, and an incident energy of $E_i=8.2$~meV, shown in Fig.~\ref{flo:FullSpectra}. Data were integrated over all directions perpendicular to $L$ \cite{Lake2005}. At each temperature, the spectra was measured for eight hours at nominal 1.4~Mw operation \cite{mason2006spallation}, with choppers set to 120 Hz (SEQ-100-2.0-AST chopper with 2 mm slit spacing). Data were corrected for the form factor by calculating the magnitude of $Q$ for each pixel and dividing by 
	the anisotropic $d_{x^2-y^2}$ Cu$^{2+}$ form factor:
	\begin{equation}
	f({\bf Q}) = \langle j_0 \rangle +
	\frac{5}{7}(3 \cos^2 \beta - 1) \langle j_2 \rangle +
	\frac{3}{56}(35 \cos^4 \beta - 30 \cos^2 \beta + 35 \sin^4 \beta \cos 4 \alpha + 3)\langle j_4 \rangle \cite{boothroyd2020principles}
	\end{equation}
	where $\beta$ is the angle between $\bf Q$ and the $z$ axis of the  $d_{x^2-y^2}$ orbital, and $\alpha$ is the $xy$-plane angle from the $x$ axis. We used Cu$^{2+}$ $\langle j_n \rangle$  constants from Ref. \cite{BrownFF}.
As shown in Fig.~\ref{flo:FF_correction}, each pixel at nonzero $Q_{L}$ (along the chain) or $\hbar\omega$ includes a nonzero $Q_{HH}$ component
\begin{equation}
Q_{HH}=\sqrt{k_f^2 - Q_{L}^2} - k_i
\end{equation}
where $k_f$ and $k_i$ are the magnitudes of the incident and final neutron wavevectors. 
	Taking this into account, we calculated the anisotropic form factor for the two $d_{x^2-y^2}$ orbital orientations shown in Fig. 1(a) in the main text, averaging over both orientations. The final calculated form factor for this geometry is in Fig.~\ref{flo:FF_correction}b.
	Fig. \ref{flo:FF_anisotropy} compares the isotropic and anisotropic Cu$^{2+}$ form factors: the difference is noticeable but small. The fitted 300~K dynamic exponent is 1.36(5) with the isotropic form factor correction, and 1.35(5) with the anisotropic form factor correction: no difference to within uncertainty.

\begin{figure}
	\centering
	\includegraphics[width=0.9\textwidth]{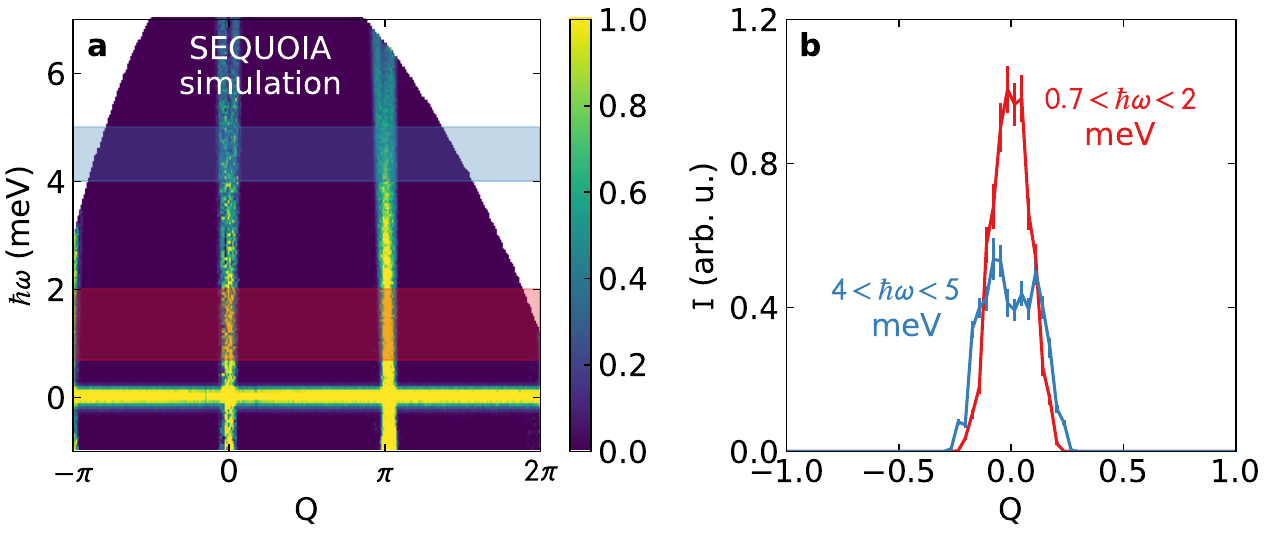}
	\caption{Simulated neutron spectrum of a $1$D antiferromagnetic spin chain in the SEQUOIA experiment, in order to show resolution width. The inelastic spectra is assumed to be the semi-classical linear spin wave theory dispersion, and the $Q$ resolution has a FWHM $0.08$. Panel \textbf{\textsf{a}} shows the spectrum, and panel \textbf{\textsf{b}} shows constant-energy cuts. At $4.5$~meV, the two modes of the dispersion are just barely distinguishable.}
	\label{flo:SimulatedSpectra}
\end{figure}

In order to estimate the experimental resolution of the SEQUOIA spectrometer for this experiment, we simulated an antiferromagnetic linear chain spin wave dispersion with a bandwidth $52.7$~meV (the lower bound of the spinon continuum) using MCViNE virtual neutron experiment \cite{lin2016mcvine}. The Monte Carlo ray tracing simulations of SEQUOIA \cite{Granroth2006}  were run using McVine for $2 \times 10^{10}$ incident neutron packets using the exact  incident energy and chopper settings used during the experiment. The results are shown in Fig.~\ref{flo:SimulatedSpectra}, and indicate a $Q$ resolution has a FWHM $0.08$. Note that the finite-temperature mode softening and the power law in $Q$ are absent from this simulated data, indicating that they are intrinsic to KCuF$_3$ and not resolution effects.

\begin{figure}
		\centering
		\includegraphics[width=0.6\textwidth]{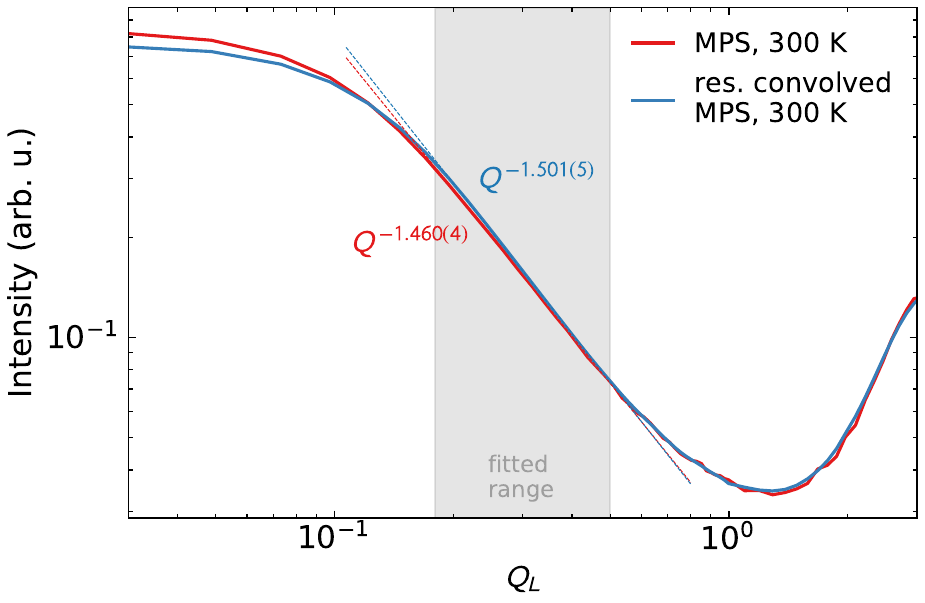}
		\caption{Effect of resolution broadening on the fitted power law, demonstrated using the MPS 300~K simulation. The resolution-broadened fit increases the fitted exponent by 2.7\%, which means that the fitted experimental data slightly overestimates the dynamic exponent.}
		\label{flo:ResolutionEffect}
	\end{figure}

	To estimate the impact of experimental resolution on the fitted power law, we used the resolution function defined by the MCViNE simulations in Fig. \ref{flo:SimulatedSpectra}, and convolved the 300~K MPS data with a Gaussian resolution, shown in Fig. \ref{flo:ResolutionEffect}. We find that the fitted exponent of the broadened data increases slightly (by 2.7\%) because of the sharp $Q=0$ feature becoming smoothed out.  This means that the experimental fits slightly overestimate the dynamic exponent: the true 300~K KCuF$_3$ dynamic exponent may be closer to 1.31(5) than 1.35(5).
	
\subsection{Data fitting}

\begin{figure}
		\centering
		\includegraphics[width=\textwidth]{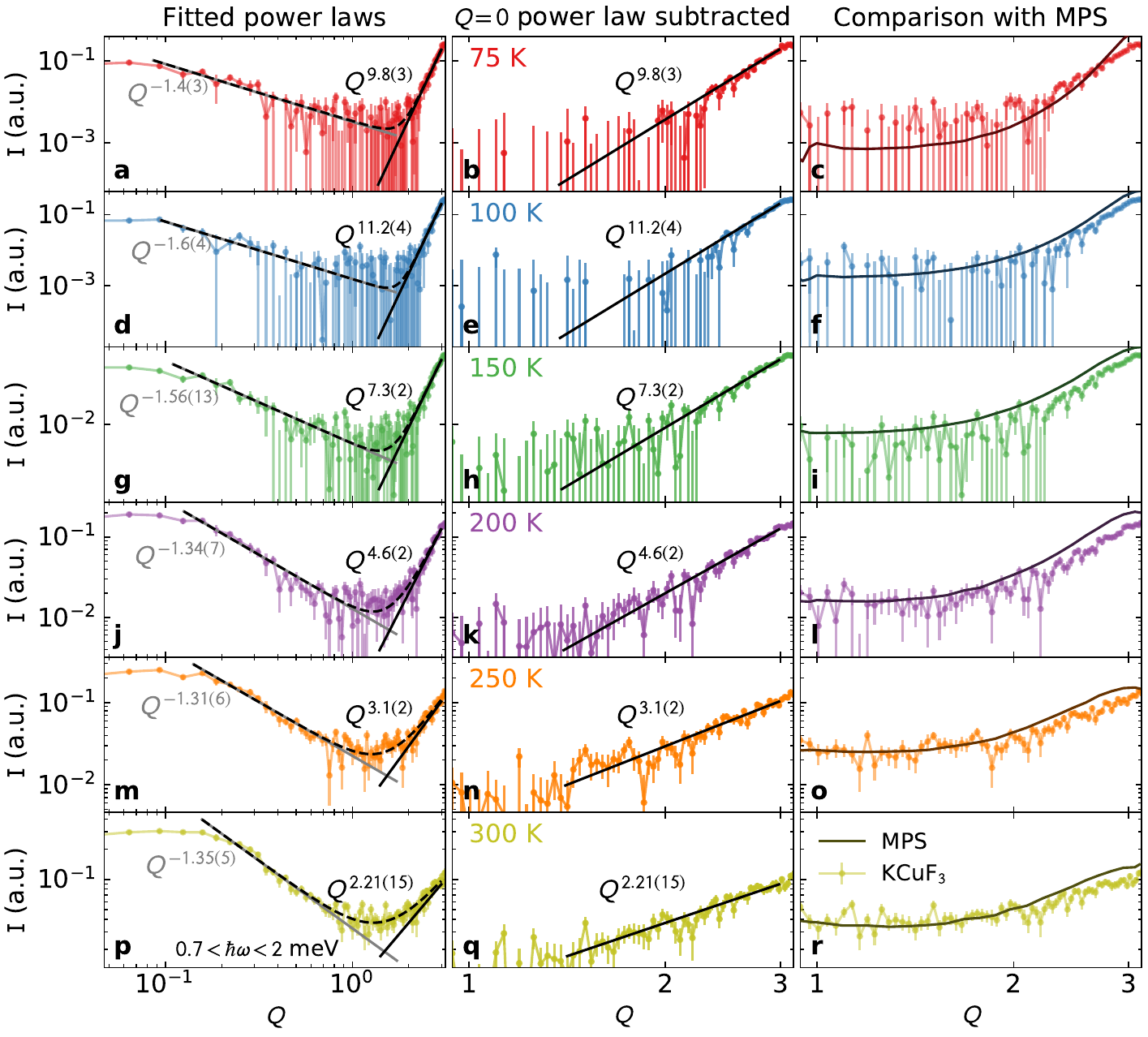}
		\caption{Power law fits for KCuF$_3$ scattering showing the fitted phenomenological power law at $Q=\pi$. The left column shows the data with the $Q=0$ and $Q=\pi$ fitted power laws, the middle column shows the $Q=\pi$ power law with the $Q = 0$ (dynamic exponent) subtracted, and the right column shows the experimental data compared to MPS simulations in the vicinity of $Q=\pi$. The data near $Q=\pi$ follows a power law very well, but it dramatically varies with temperature. The right column shows this is the case for both theory and experiment.}
		\label{flo:Q1_power_law}
	\end{figure}

\begin{figure}
	\centering
	\includegraphics[width=\textwidth]{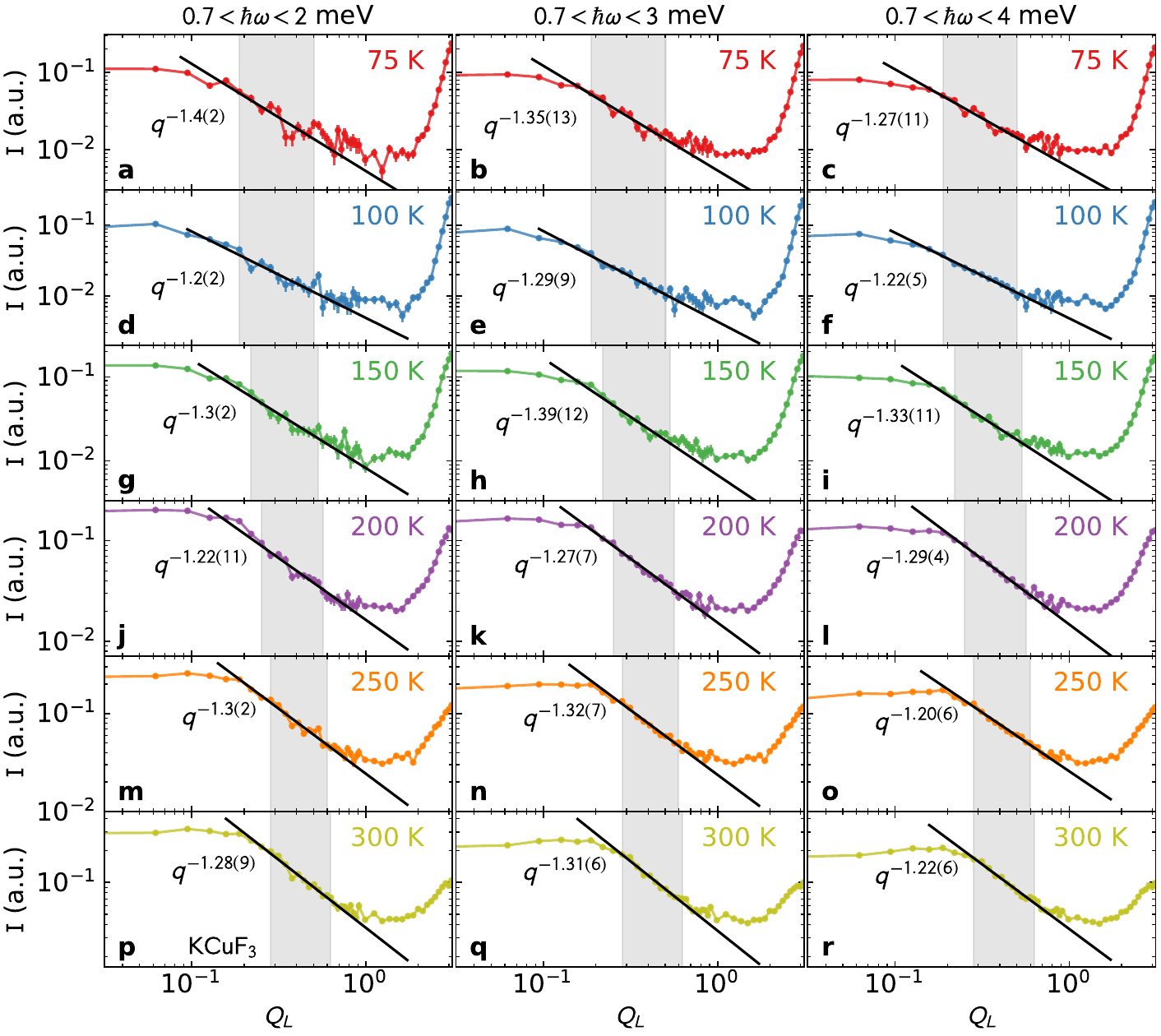}
	\caption{Power law fits for low-$Q$ KCuF$_3$ scattering for three different energy windows: $0.7<\hbar \omega<2$~meV (cut a, left column), $0.7<\hbar\omega<3$~meV (cut b, middle column), and $0.7<\hbar\omega 4$~meV (cut c, right column). No background has been subtracted, and the window where data was fitted is indicated in gray. Note that these powers, in most cases, agree to within uncertainty with those in Fig.~$4$ in the main text.}
	\label{flo:data_PLFits}
\end{figure}

In Fig.~$3$ of the main text, we subtract a phenomenological power law at $Q=\pi$ in order to isolate the power law at $Q=0$. 
	The fitted power laws are shown in Fig. \ref{flo:Q1_power_law}, showing that the power near $Q=\pi$ dramatically varies with temperature both in theory and experiment. 
We also chose the lowest energy window (``cut a'' in Fig.~2 of the main text) to approximate the $\hbar\omega=0$ scattering. If we do not do this, but instead define a constant background based off the lowest temperature data, the fitted exponents are shown in Fig.~\ref{flo:data_PLFits}. We also show the results of fits to cuts b and c. 
Forgoing the phenomenological background means that the power law is visible over a narrower range in $Q$, but in nearly every case the fitted powers agree to within uncertainty. Meanwhile, an increase in the energy window yields slightly different fitted powers, generally decreasing as the window increases.
All fits were performed with the scipy least squares routine \cite{virtanen2020scipy}.

\clearpage\newpage

\section{Numerical simulations}
\subsection{Method}
\subsubsection{Finite temperature}

The numerical simulations are based on a matrix product state (MPS) method~\cite{schollwock2011} using the ITensor library~\cite{itensor} to simulate the finite-temperature spin dynamics of the $1$D quantum Heisenberg spin-half chain, see Eq.~(1) of the main text (the antiferromagnetic exchange coupling is $J_c=33.5$~meV). We use the purification method to represent the finite-temperature quantum state $\hat{\rho}_T$ (mixed state) as a pure state $|\psi_T\rangle$ in an enlarged Hilbert space~\cite{verstraete2004}. In practice, this enlarged Hilbert space corresponds to doubling the system size (from $L$ to $2L$ degrees of freedom) with half physical ``$\mathsf{P}$'' and half auxiliary ``$\mathsf{Q}$'' degrees of freedom and the property that $\hat{\rho}_T=\mathrm{tr}_\mathsf{Q}|\psi_T\rangle\langle\psi_T|$, where $\mathrm{tr}_\mathsf{Q}$ is the partial trace over the auxiliary spins (also called ancilla). In this approach, the thermal expectation value of an observable $\hat{\mathcal{O}}$ at temperature $T$ is expressed as,
\begin{equation}
\left\langle\hat{\mathcal{O}}\right\rangle= \Bigl\langle\psi_T\Bigl|\hat{\mathcal{O}}\Bigr|\psi_T\Bigr\rangle\Bigl/\Bigl\langle\psi_T\Bigl|\psi_T\Bigr\rangle,
\end{equation}
with
\begin{equation}
|\psi_T\rangle=\exp\left(-\frac{\hat{\mathcal{H}}}{2k_\mathrm{B}T}\right)|\psi_\infty\rangle. \label{eq:psiT}
\end{equation}
Here, $\hat{\mathcal{H}}=\hat{\mathcal{H}}_\mathsf{P}\otimes\hat{I}_\mathsf{Q}$ with $\hat{\mathcal{H}}_\mathsf{P}$ the Hamiltonian of the physical system (and which only acts on the physical spins). The state $|\psi_{\infty}\rangle$ has an exact MPS representation and can be constructed exactly: it is a product state of $L$ pairs, each involving one physical and one auxiliary spin. Each pair corresponds to a maximally entangled state (typically a singlet). In the following, in order to lighten notations, we will omit the tensor product with $\hat{I}_\mathsf{Q}$ when referring to operators acting on physical degrees of freedom, unless specified otherwise. The Heisenberg Hamiltonian of Eq.~(1) of the main text being local with only nearest neighbor interaction terms, we perform the imaginary-time evolution of Eq.~\eqref{eq:psiT} using the time-evolving block decimation algorithm~\cite{Vidal2004} along with a fourth order Suzuki-Trotter decomposition~\cite{Garcia2006}. Typically, we use time step $\tau=0.1$, or a close value commensurate with the desired inverse temperature and the number of Trotter steps $N_\tau$ such that $\tau N_\tau=1/2k_\mathrm{B}T$.

\subsubsection{Dynamics and spectral function}

When the state $|\psi_T\rangle$ is obtained and normalized such that $\langle\psi_T|\psi_T\rangle=1$ (this corresponds to setting the partition function of the system to $1$), we use a method expanding the desired spectral function $\mathcal{S}(Q,\omega)$ in terms of Chebyshev polynomials~\cite{holzner2011, AWolf2015}. The dynamics is generated by the Louivillian operator $\hat{\mathcal{L}}=\hat{\mathcal{H}}_\mathsf{P}\otimes\hat{I}_\mathsf{Q} - \hat{I}_\mathsf{P}\otimes\hat{\mathcal{H}}_\mathsf{Q}$, i.e., the dynamical structure factor is expressed directly in frequency space as~\cite{Barnett1987,tiegel2014},
\begin{equation}
\mathcal{S}\bigl(Q,\omega\bigr)=\Bigl\langle\hat{\boldsymbol{S}}_{-Q}\delta\Bigl(\hbar\omega-\hat{\mathcal{L}}\Bigr)\cdot\hat{\boldsymbol{S}}_Q\Bigr\rangle,
\label{eq:Sqw}
\end{equation}
with the momentum space spin operators defined by,
\begin{equation}
\hat{\boldsymbol{S}}_Q=\sqrt{\frac{2}{L+1}}\sum_{r=1}^L\sin\bigl(Qr\bigr)\hat{\boldsymbol{S}}_r,
\end{equation}
with $L$ the total number of spins in the system with lattice spacing taken equal to unity; $r$ labels the spin index on the chain. Because MPS are more efficient at simulating systems with open boundary conditions, we have used a slightly different version of the Fourier transform in the above equation compared to the usual definition~\cite{Benthien2004}; both are equivalent in the thermodynamic limit $L\to+\infty$. Here, the allowed momentum by the finite-length geometry are $Q=k\pi/(L+1)$ with $k=1,2,\,\ldots,\,L$. Because the Heisenberg model is isotropic with respect to the different spin components, we can make the substitution $\bigl\langle\hat{\boldsymbol{S}}_{-Q}\delta\bigl(\omega-\hat{\mathcal{L}}\bigr)\cdot\hat{\boldsymbol{S}}_Q\bigr\rangle\to 3\bigl\langle\hat{S}^\alpha_{-Q}\delta\bigl(\omega-\hat{\mathcal{L}}\bigr)\hat{S}^\alpha_Q\bigr\rangle$ and only compute the dynamics associated with the $\alpha\in [x,\,y,\,z]$ spin component. In practice we choose $\alpha\equiv z$, and ignore the factor of $3$ as the overall scale is adjusted to compare with experiments whose spectral intensity is in arbitrary units (a.u.).

To compute the spectral function numerically, we expand the delta function of Eq.~\eqref{eq:Sqw} in a Chebyshev series~\cite{holzner2011,AWolf2015}. An expansion in Chebyshev polynomials, $T_n(x)=\cos\bigl[n\arccos(x)\bigr]$, is only permitted for $x\in[-1,1]$. Thus, we rescale the spectrum of $\hat{\mathcal{L}}$ and $\omega$ by an amount $W$ to ensure that the spectral function $\mathcal{S}(Q,\omega)$ is only non-zero in the range $\omega\in[-1+\epsilon,1-\epsilon]$. Using a small $\epsilon>0$ helps with numerical stability. The rescaled quantities are defined by,
\begin{equation}
\hat{\mathcal{L}}'=\hat{\mathcal{L}}\bigl/W,\quad\omega'=\omega\bigl/W.
\end{equation}
Then, the dynamical structure factor is written
\begin{equation}
\mathcal{S}\bigl(Q,\omega\bigr)= \frac{1}{W\pi\sqrt{1-\omega'^2}}\sum_{n=0}^{\infty}\bigl(2-\delta_{n0}\bigr)\mu_n\bigl(Q\bigr)T_n\bigl(\omega'\bigr),
\label{eq:Sqw_Cheb}
\end{equation}
with $\mu_n(Q)=\langle t_0|t_n\rangle$ the Chebyshev moments and $|t_n\rangle=T_n\bigl(\hat{\mathcal{L}}'\bigr)\hat{S}_Q^z|\psi_T\rangle$ the Chebyshev vectors. They can be calculated iteratively by
\begin{equation}
|t_n\rangle=2\hat{\mathcal{L}}'|t_{n-1}\rangle-|t_{n-2}\rangle,
\label{eq:iteration}
\end{equation}
with $|t_0\rangle=\hat{S}_Q^z|\psi_T\rangle$ and $|t_1\rangle=\hat{\mathcal{L}}'|t_0\rangle$ as a starting point. For comparison between experiments and numerical simulations, we use a system size of $L=128$ spins, a bond dimension $\chi=256$, and the order of the Chebyshev expansion is $N=3000$. We chose $W=L/2(1-\epsilon)$ together with $\epsilon=0.0125$ for this work.

\paragraph{Gaussian broadening.---} Extracting an analytic spectral function from simulations is difficult due to the discrete spectrum for any finite system. A common technique is to broaden the delta functions with a smooth distribution, such as a Gaussian. Here, this is achieved by scaling the moments, $\mu_n(Q)$, by a damping factor $g_n$ that also smooths out the Gibbs oscillations that occur from truncating the series in Eq.~\eqref{eq:Sqw_Cheb} to a finite value $N$. In this work, we occasionally use Jackson damping~\cite{weibe2006}, with
\begin{align}
g_n = \frac{N-n+1}{N+1}\cos\left(\frac{\pi n}{N+1}\right) + \frac{1}{N+1}\sin\left(\frac{\pi n}{N+1}\right)\cot\left(\frac{\pi}{N+1}\right).
\end{align}
Note that $g_n$ is a monotonic function of $n$ that decays from $1$ to $0$, and in the limit $N\to+\infty$, $g_n\to1$ for all $n$. Jackson damping has the effect of smoothing out the finite system spectral peaks with a Gaussian with an $\omega$ dependent width. This Gaussian broadening procedure is used in Figs.~\ref{fig:color-low} and~\ref{fig:color-high}, as well as in Fig.~4 of the main text.

\begin{figure}
	\centering
	\includegraphics[width=0.5\linewidth]{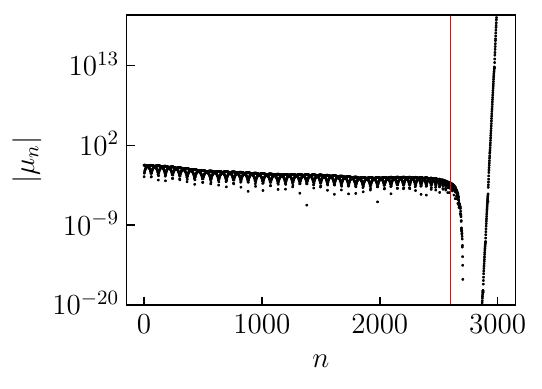}
	\caption{The absolute value of the Chebyshev moments, $|\mu_n|$, as a function of $n$ for the case of $T=75$~K, and $Q=1.680$. The vertical line in red illustrates $N^{*}$, the value of $n$ for which we terminate the series in Eq.~\eqref{eq:Sqw_Cheb}.}
	\label{fig:trim}
\end{figure}

\paragraph{Chebyshev moments.---} The Chebyshev moments, $\mu_n(Q)$, are the expansion coefficients in Eq.~\eqref{eq:Sqw_Cheb}, and dictate the convergence of the numerical simulations. The moments have an envelope that decays exponentially~\cite{AWolf2015}, and if $\mu_n(Q)$ is not ``sufficiently close to zero'', then our simulations are unreliable. We fit the envelope of the moments to a decaying exponential of the form $\propto e^{-n/\xi}$, and if $\xi>500$, we say the moments are not converged. This procedure defines a minimum $Q$ value, $Q_\mathrm{min}$, and we don't show $Q<Q_\mathrm{min}$ when comparing with experiments in the main text. We will discuss this more in Sec.~\ref{sec:N}, and show the moments in Fig.~\ref{fig:qmin-high} and Fig.~\ref{fig:qmin-low}. Numerical error can spoil the iterative process of Eq.~\eqref{eq:iteration}, and can cause $\mu_n$ to rapidly drop to near zero, and then diverge for values of $n>N^{*}$. When this happens, we zero out the moments for all $n>N^{*}$ when computing $\mathcal{S}(Q,\omega)$ in the sum of Eq.\eqref{eq:Sqw_Cheb}, effectively truncating the series at $N^{*}$. This occurs only at $T=100$ K, and $T=75$ K, and for $Q$ values larger than the power law region. The lowest value of $N^{*}$ used is $2600$, and we show in Fig.~\ref{fig:trim} the moments for this case.

\subsection{Additional numerical data}
\subsubsection{Spectral function}

\begin{figure}
	\centering
	\includegraphics[width=0.7\linewidth]{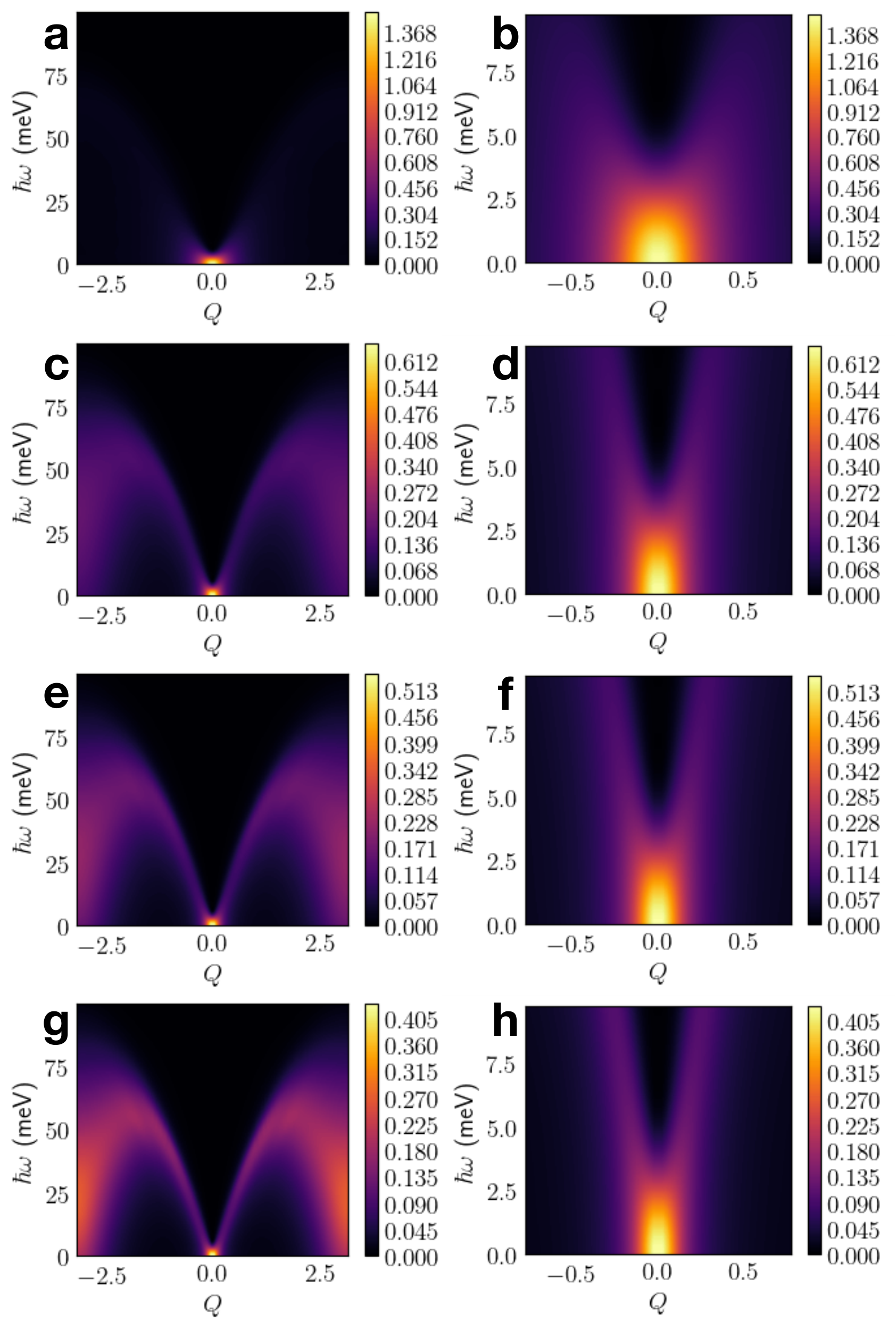}
	\caption{The spectral function, $\mathcal{S}(Q,\omega)$ at \textbf{\textsf{a}}, \textbf{\textsf{b}} $T=+\infty$, \textbf{\textsf{c}}, \textbf{\textsf{d}} $300$~K, \textbf{\textsf{e}}, \textbf{\textsf{f}} $250$~K, and \textbf{\textsf{g}}, \textbf{\textsf{h}} $200$~K. The left column illustrates the full spectrum, and the right column zooms in around $(Q,\omega)=(0,0)$, the region relevant for comparison with experiments.}
	\label{fig:color-high}
\end{figure}

\begin{figure}
	\centering
	\includegraphics[width=0.7\linewidth]{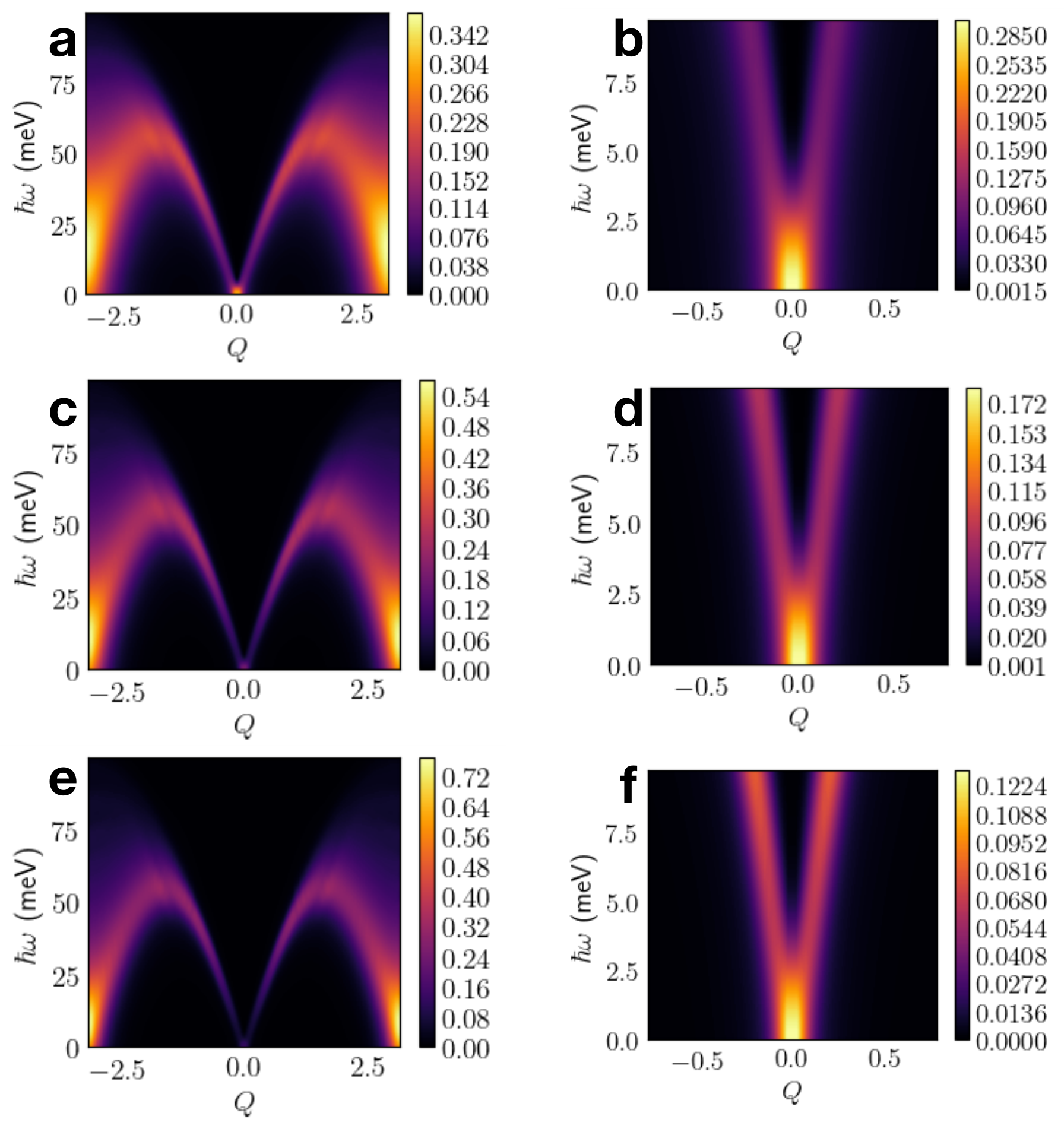}
	\caption{The spectral function, $\mathcal{S}(Q,\omega)$ at \textbf{\textsf{a}}, \textbf{\textsf{b}} $T=150$~K, \textbf{\textsf{c}}, \textbf{\textsf{d}} $100$~K, and \textbf{\textsf{e}}, \textbf{\textsf{f}} $75$~K. The left column illustrates the full spectrum, and the right column zooms in around $(Q,\omega)=(0,0)$, the region relevant for comparison with experiments.}
	\label{fig:color-low}
\end{figure}

\begin{figure}
	\centering
	\includegraphics[width=0.7\linewidth]{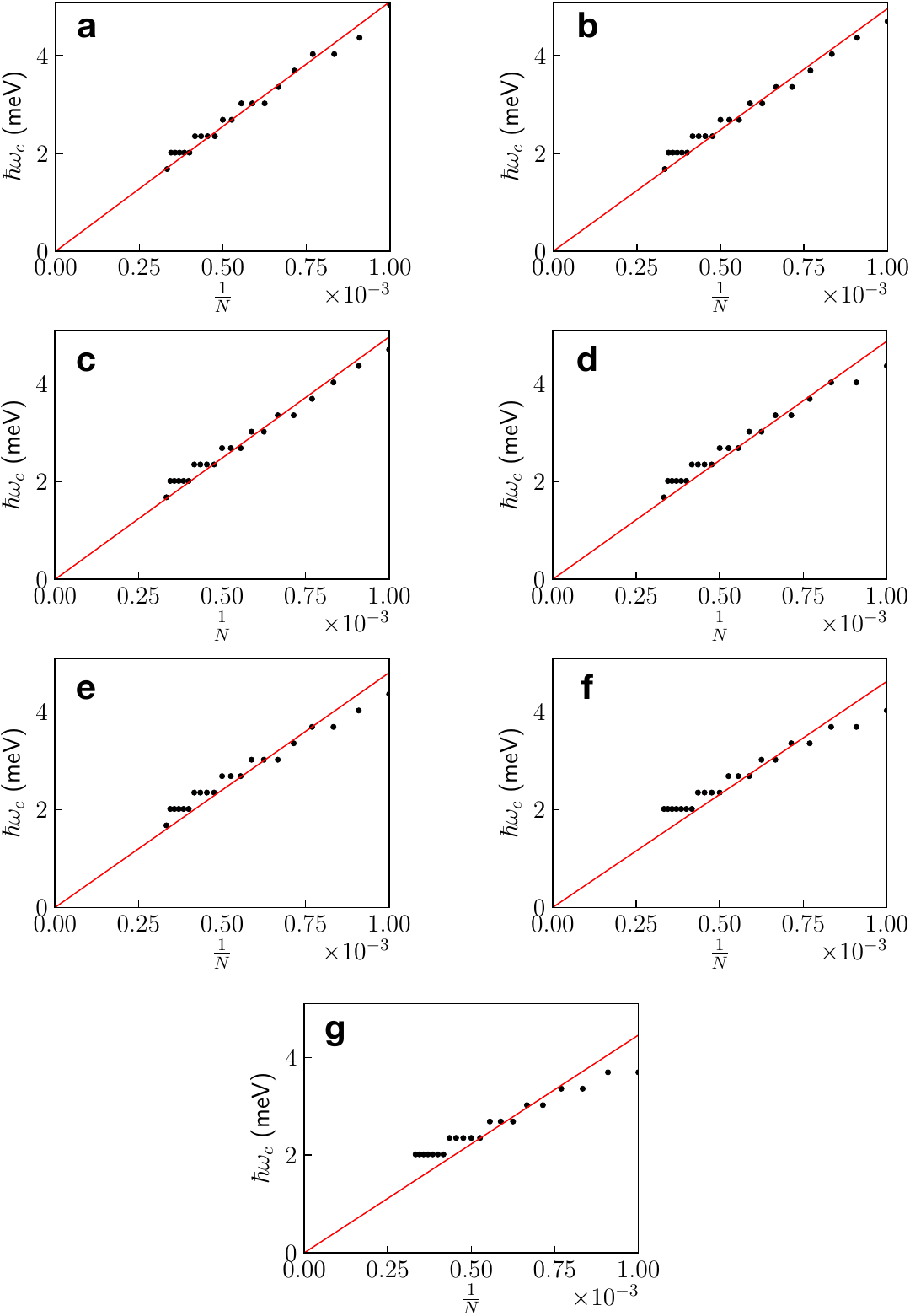}
	\caption{The value of $\hbar\omega_\mathrm{c}$ at which the spinon mode splitting occurs at $Q=0$ versus the inverse of the number of terms used in the sum of Eq.~\eqref{eq:Sqw_Cheb}. The different subplots correspond to \textbf{\textsf{a}} $T=+\infty$, \textbf{\textsf{b}} $300$~K, \textbf{\textsf{c}} $250$~K, \textbf{\textsf{d}} $200$~K, \textbf{\textsf{e}} $150$~K, \textbf{\textsf{f}} $100$~K, and \textbf{\textsf{g}} $75$~K.}
	\label{fig:bifurcation}
\end{figure}

The full $\omega$ dependence is computed, and shown in Fig.~\ref{fig:color-high} and Fig.~\ref{fig:color-low} for various temperature values. Jackson damping is used in these figures. As temperature is lowered, agreement with the low-energy spectrum is observed. The region near $(Q,\omega)=(0,0)$ relevant for the neutron scattering experiments is also shown. We see strong intensity at $(Q,\omega)=(0,0)$ with bifurcating dispersion lines emerging as we move away from this point. We note that the bifurcation point appears to occur at finite $\omega$ from the numerical data, but this is a finite size effect, and does not seem to occur in the experimental data. To verify this, we show the frequency bifurcation, $\omega_\mathrm{c}$ as a function of the Chebyshev expansion order $N$ in Fig.~\ref{fig:bifurcation}. As $N\to+\infty$, we're approaching the thermodynamic limit, and we see this bifurcation point tends towards $\omega_\mathrm{c}=0$.

\subsubsection{Kardar-Parisi-Zhang scaling function}

\begin{figure}
	\centering
	\includegraphics[width=0.5\linewidth]{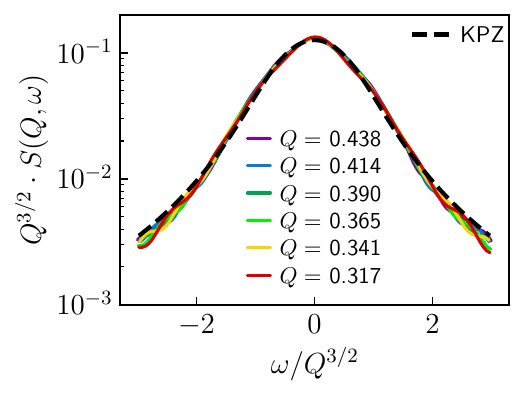}
	\caption{The KPZ scaling function plotted against the numerical data for various values of $Q$ at $T=+\infty$.}
	\label{fig:scaling}
\end{figure}

The scaling function for the $1+1$ KPZ universality class $f(\cdot)$ is known numerically exactly~\cite{prahofer2004} in real space $x$ and time $t$. In particular, Ref.~\cite{prahofer2004} provides raw data for the scaling function $f(y)$ with $y\propto xt^{-2/3}$, and following this work, we Fourier transform this data to arrive at the scaling function $\mathring{f}(\cdot)$ in momentum $Q$ and frequency $\omega$ space. The relation to the spectral function, in the hydrodynamic regime is given by
\begin{equation}
\mathcal{S}(Q,\omega)=c_1Q^{-3/2}\mathring{f}\Bigl(c_2\omega Q^{-3/2}\Bigr),
\label{eq:scaling}
\end{equation}
where $c_1$ and $c_2$ are system-dependent constants that we use as fitting parameters to compare with the numerical simulations. We found $c_1\simeq0.026$ and $c_2\simeq0.639$ by fitting $\mathring{f}$ to the numerical data at $T=+\infty$ and $Q=Q_\mathrm{min}$ for the optimum simulation parameters used in this work. Comparison with our numerical simulations and the scaling functions for KPZ are shown in Fig.~\ref{fig:scaling}.

\subsubsection{Robustness of the power law fit with the energy window}

\begin{figure}
	\centering
	\includegraphics[width=\textwidth]{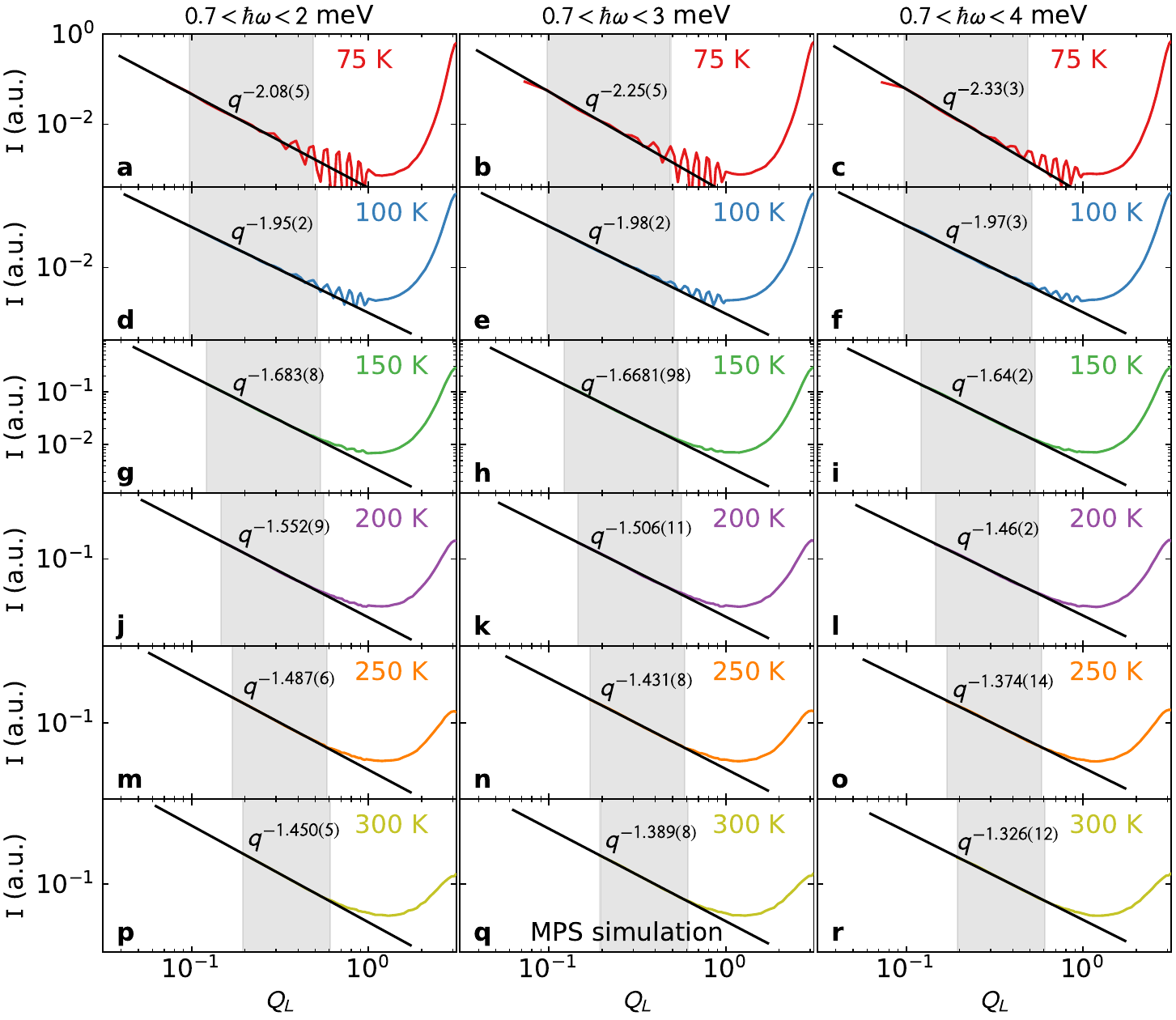}
	\caption{Power law fits to the low-$Q$ MPS simulated spectrum for three different energy windows: $0.7<\hbar\omega<2$~meV (cut a, left column), $0.7<\hbar\omega<3$~meV (cut b, middle column), and $0.7<\hbar\omega<4$~meV (cut c, right column). The window where data was fitted is indicated in gray. Note that the finite energy transfer causes the exponent to deviate from the $\hbar\omega=0$ value of $z=3/2$: too large at low temperatures, and too high at high temperatures. Also note that the exponent magnitude is suppressed as the energy transfer window increases.}
	\label{flo:mps_PLFits}
\end{figure}

We show the power law fit for different choices of the energy window at the temperatures relevant for experiments in Fig.~\ref{flo:mps_PLFits}, analogous to Fig.~\ref{flo:data_PLFits}. In this case, the trends are more clear: at lower temperatures, the fitted power is larger, and larger energy windows suppresses the fitted power at high temperatures, and enhances the fitted power at the lowest temperatures.

\begin{figure}
	\centering
	\includegraphics[width=0.6\textwidth]{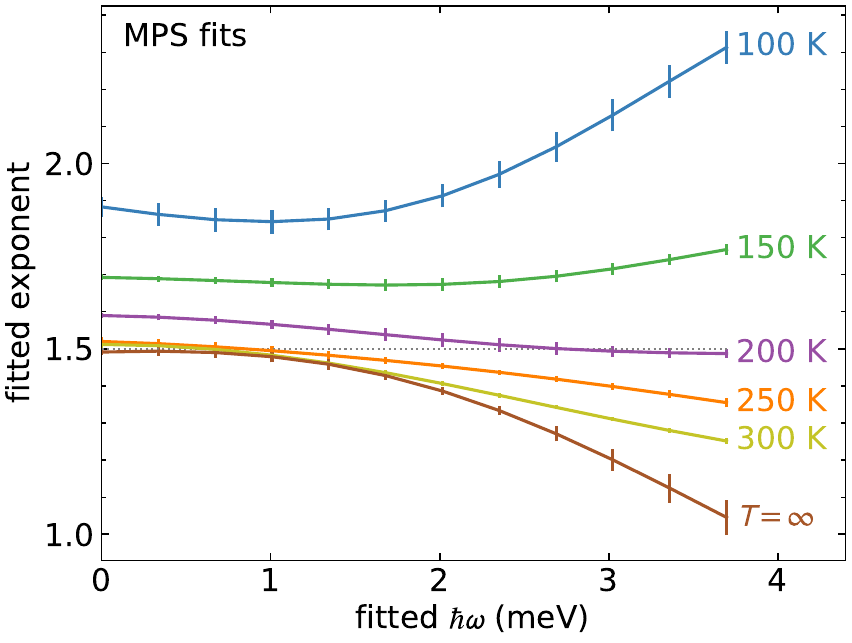}
	\caption{Deviations from $Q^{-3/2}$ KPZ behavior at finite temperature and finite energy transfer. Each curve shows the fitted exponent for the MPS simulated low-$Q$ scattering as a function of energy. At infinite temperature and $\hbar\omega=0$, the exponent is $-3/2$. As temperature decreases, the exponent generally increases. However, above $200$~K, the KPZ behavior is still dominant.}
	\label{flo:mps_PL_deviations}
\end{figure}

These effects are shown more systematically in Fig.~\ref{flo:mps_PL_deviations}, which show deviations from the ideal KPZ $z=3/2$ exponent as temperature becomes finite and the fitted energy increases. At infinite temperature and zero energy, the fitted exponent is almost exactly $-3/2$. At 300 K and at low energy, the fitted exponent is still quite close. However, as temperature drops the $\hbar\omega$ fitted exponent drifts toward $-2$, while the finite-energy fitted value ranges from below $1$ to above $2$, depending on the temperature. These effects are important to take into account, because any real experiment will measure the exponent with a finite temperature and non-zero energy transfer.

\subsection{Convergence checks with simulation parameters}
\subsubsection{Bond dimension of the matrix product state}

\begin{figure}
	\centering
	\includegraphics[width=\linewidth]{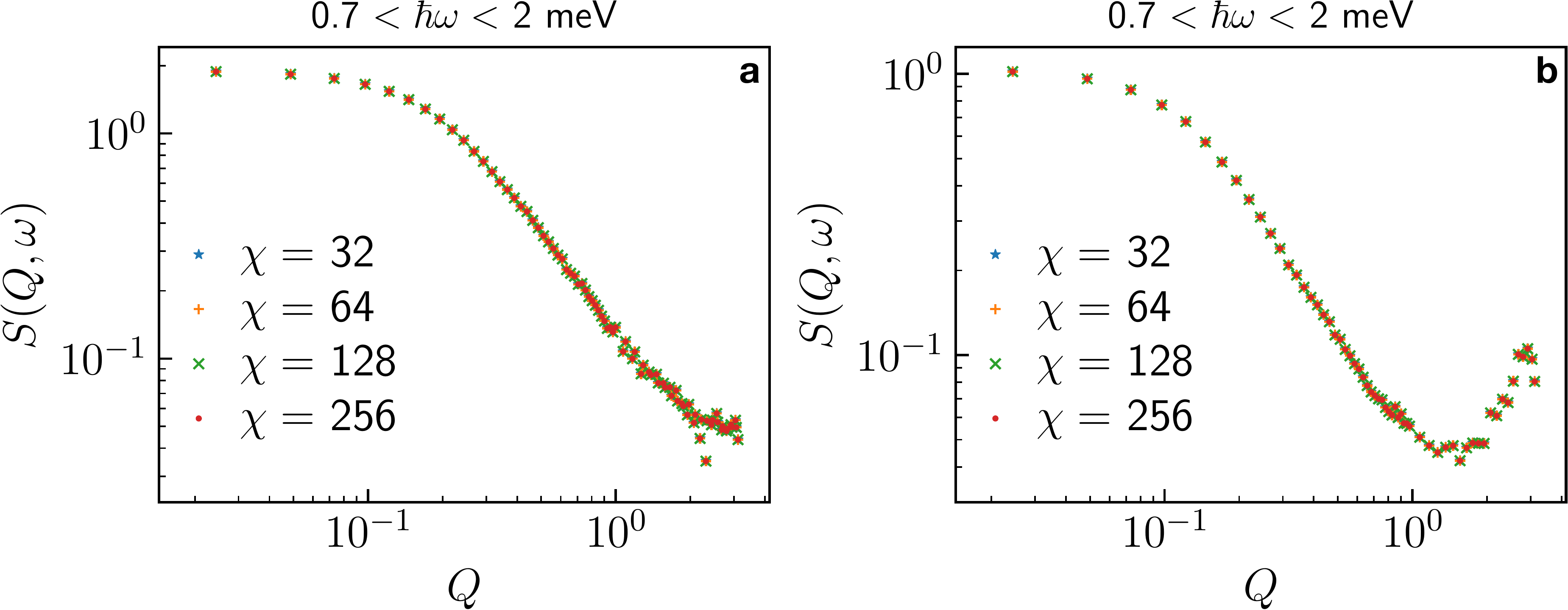}
	\caption{The $Q$ dependence of $\mathcal{S}(Q,\omega)$ for $0.7<\hbar\omega<2$~meV for different choice of the MPS bond dimension $\chi$. Subplot a corresponds to $T=+\infty$, and b corresponds to $T=J_c\simeq 390$~K.}
	\label{fig:sqw0chi}
\end{figure}

\begin{figure}
	\centering
	\includegraphics[width=\linewidth]{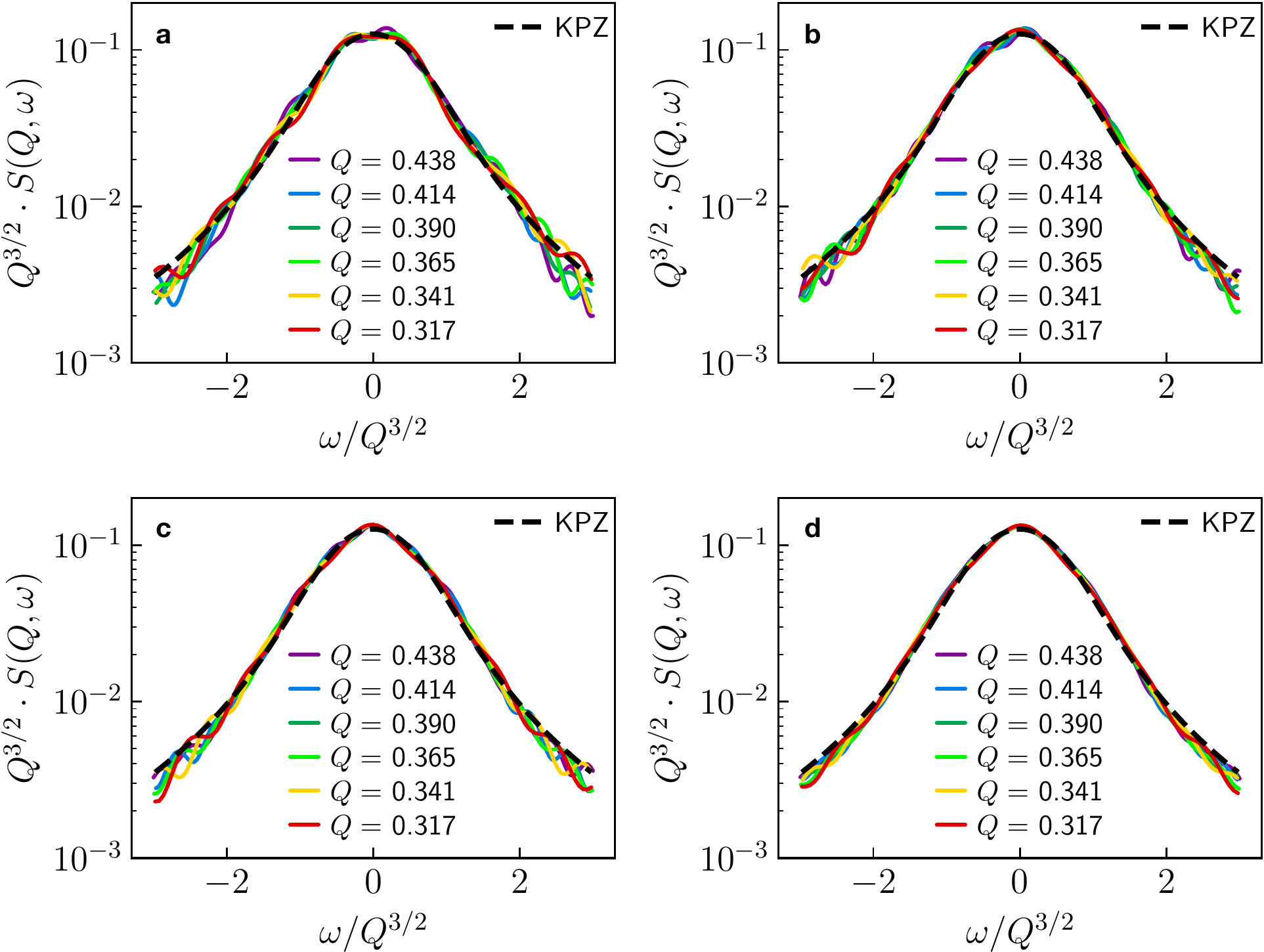}
	\caption{The KPZ scaling function plotted against the numerical data for various values of $Q$ at $T=+\infty$. The different subplots correspond to \textbf{\textsf{a}} $\chi=32$, \textbf{\textsf{b}} $\chi=64$, \textbf{\textsf{c}} $\chi=128$, and \textbf{\textsf{d}} $\chi=256$.}
	\label{fig:scalingchi}
\end{figure}

For comparison with experiments, we used a bond dimension $\chi=256$, and here we show the effects of changing $\chi$. When examining the $Q$ dependence for low $\omega$, the effect of $\chi$ is insignificant as illustrated in Fig.~\ref{fig:sqw0chi}, suggesting convergence, at least in the $Q$ and $\omega$ regime of interest. In this figure, we integrate over the window $0.7<\hbar\omega<2$~meV, which corresponds with the energy window used in comparison with experiments. Lastly, we show the effect of $\chi$ on the scaling function in Fig.~\ref{fig:scalingchi}, where we see the oscillations decay with increasing $\chi$, but overall good agreement with the KPZ scaling function is observed.

\subsubsection{System size and finite size effects}

\begin{figure}
	\centering
	\includegraphics[width=\linewidth]{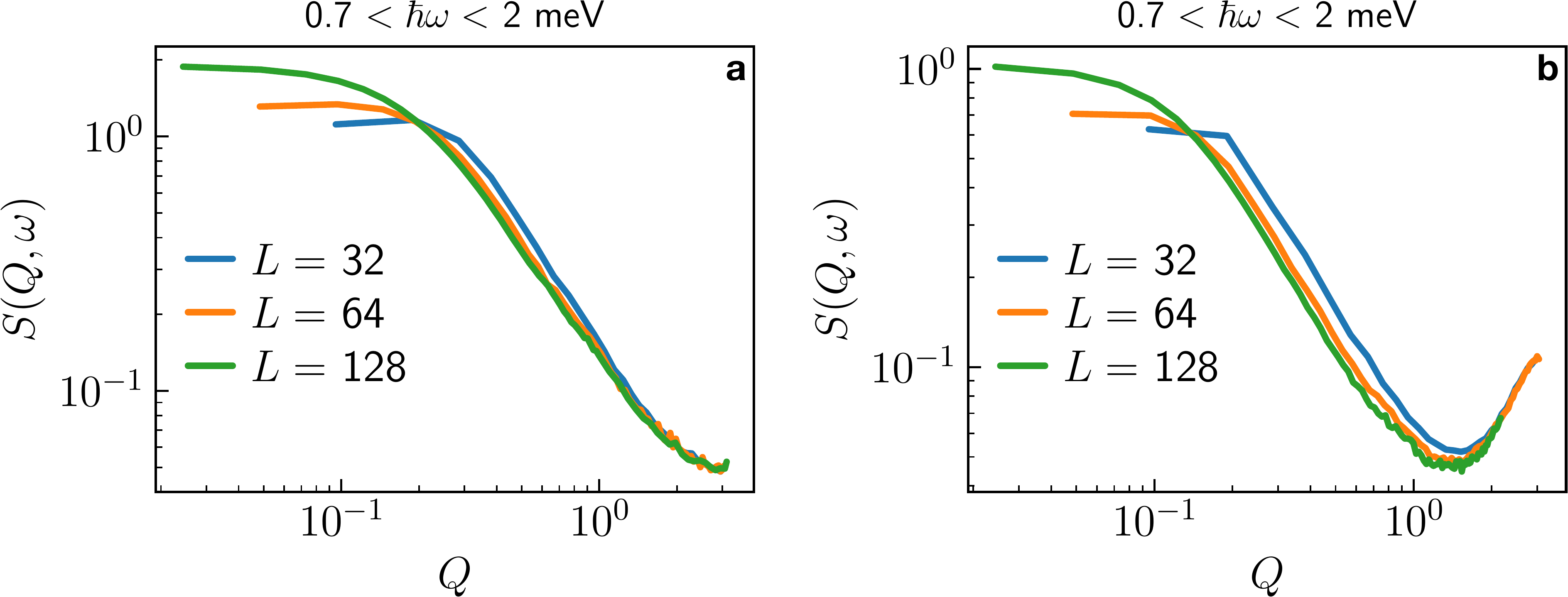}
	\caption{The $Q$ dependence of $\mathcal{S}(Q,\omega)$ for $0.7\hbar\omega<2$~meV with different choice of the system size $L$. Subplot \textbf{\textsf{a}} corresponds to $T=+\infty$, and \textbf{\textsf{b}} corresponds to $T=J_c\simeq 390$~K.}
	\label{fig:sqw0L}
\end{figure}

\begin{figure}
	\centering
	\includegraphics[width=\linewidth]{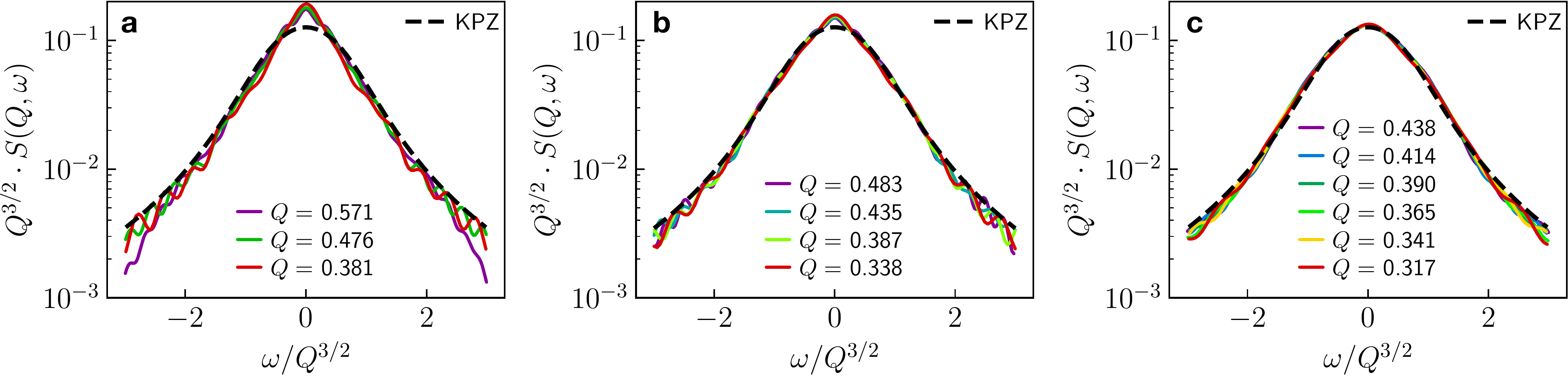}
	\caption{The KPZ scaling function plotted against the numerical data for various values of $Q$ at $T=+\infty$. The different subplots correspond to \textbf{\textsf{a}} $L=32$, \textbf{\textsf{b}} $L=64$, and \textbf{\textsf{c}} $L=128$.}
	\label{fig:scalingL}
\end{figure}

The system size $L$ is the simulation parameter that has the largest effect on the simulations. Determining the $\omega$ dependence in the thermodynamic limit from a finite system is inherently difficult as for finite $L$, $\mathcal{S}(Q,\omega)$ is a sum of delta functions, yet in the thermodynamic limit, $\mathcal{S}(Q,\omega)$ is an analytic function of $\omega$. We show the $Q$ dependence $\mathcal{S}(Q,\omega)$, integrated over $\omega$ in the window $0.7<\hbar\omega<2$~meV, in Fig.\ref{fig:sqw0L}. We see that the major difference between different system sizes is an overall multiplicative factor. Since the experimental data is collected with arbitrary units (a.u.), such an overall multiplicative factor is irrelevant when making a comparison between experiments and numerical simulations. Moreover, the curves at different system size exhibit nearly identical power law behavior for intermediate values of $Q$ and thus the dynamical exponent $z$ is quite robust to system size, and this is where the KPZ signature lies. We examine how the scaling function changes with system size in Fig.~\ref{fig:scalingL}. We notice the major difference between system size occurs for low $\omega/Q^{3/2}$, which is due to an inability to access small frequencies for small system sizes (the system size provides an artificial cutoff).

\subsubsection{Chebyshev expansion order}
\label{sec:N}

\begin{figure}
	\centering
	\includegraphics[width=0.7\linewidth]{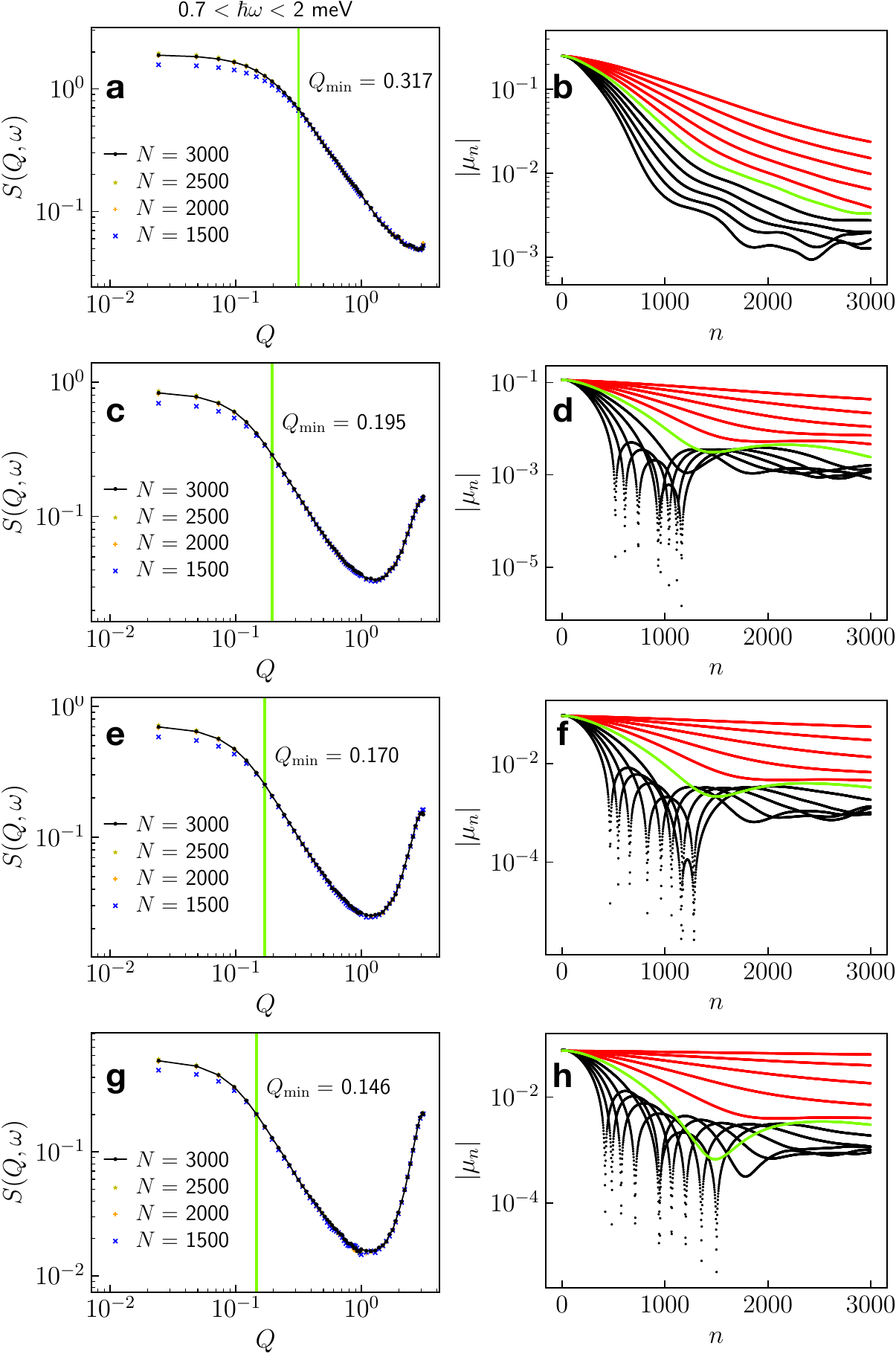}
	\caption{The left columns shows the $Q$ dependence of $\mathcal{S}(Q,\omega)$ for $0.7\hbar\omega<2$~meV with different choice of the number of terms, $N$, used in Eq.~\eqref{eq:Sqw_Cheb}. The vertical green line corresponds to the value $Q=Q_\mathrm{min}$. The right column is the absolute value of the Chebyshev moments appearing in Eq.~\eqref{eq:Sqw_Cheb} as a function of the number of iterations $n$ for several values of $Q$. Only the moments with even values of $n$ are shown for clarity. The green curve in each plot depicts $Q=Q_\mathrm{min}$, the red curves are $Q<Q_\mathrm{min}$, and the black curves are $Q>Q_\mathrm{min}$. The temperatures shown are \textbf{\textsf{a}}, \textbf{\textsf{b}} $T=+\infty$, \textbf{\textsf{c}}, \textbf{\textsf{d}} $300$~K, \textbf{\textsf{e}}, \textbf{\textsf{f}} $250$~K, and \textbf{\textsf{g}}, \textbf{\textsf{h}} $200$~K.}
	\label{fig:qmin-high}
\end{figure}

\begin{figure}
	\centering
	\includegraphics[width=0.7\linewidth]{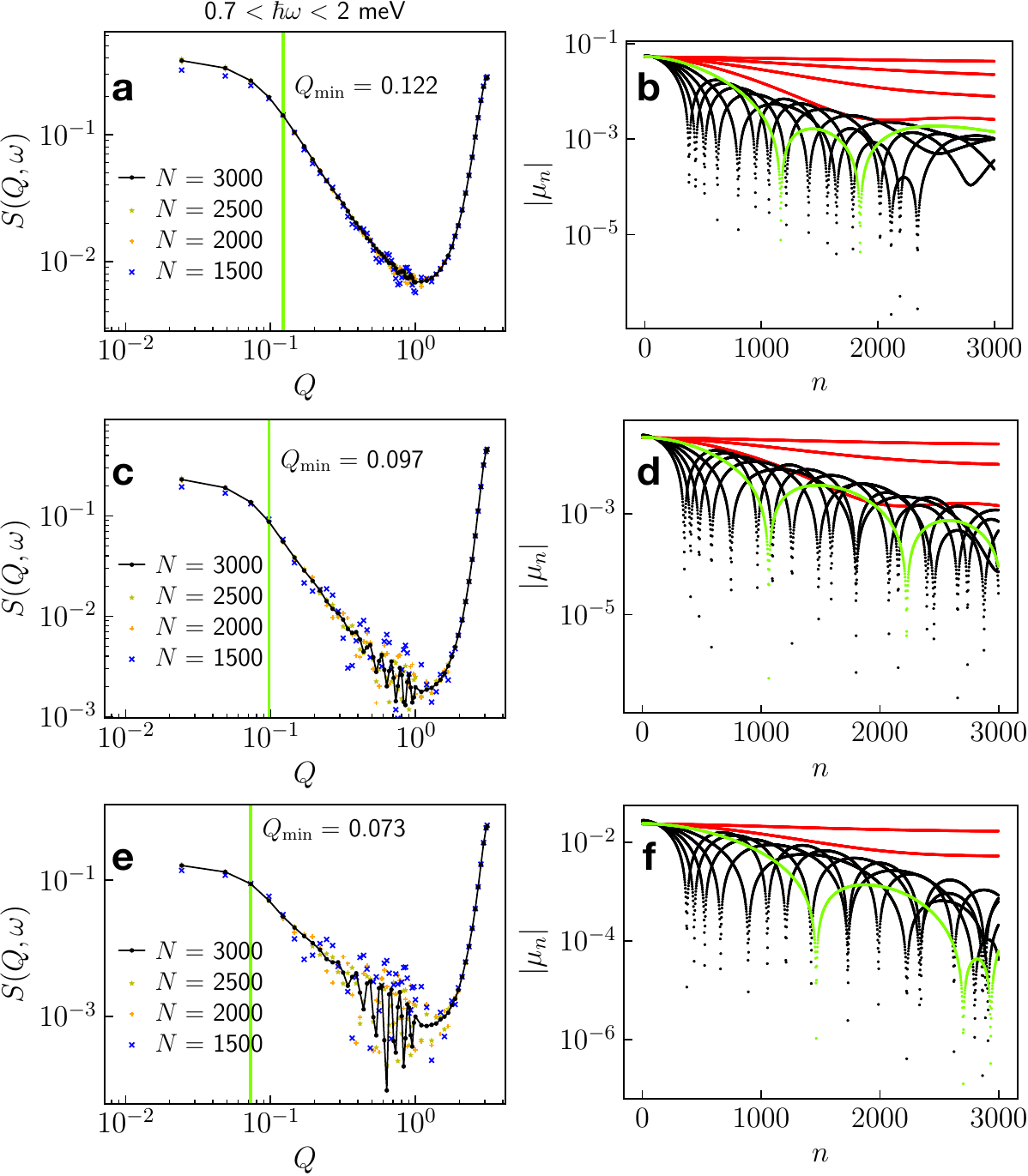}
	\caption{The left columns shows the $Q$ dependence of $\mathcal{S}(Q,\omega)$ for $0.7<\hbar\omega<2$~meV with different choice of the number of terms, $N$, used in Eq.~\eqref{eq:Sqw_Cheb}. The vertical green line corresponds to the value $Q=Q_\mathrm{min}$. The right column is the absolute value of the Chebyshev moments appearing in Eq.~\eqref{eq:Sqw_Cheb} as a function of the number of iterations $n$ for several values of $Q$. Only the moments with even values of $n$ are shown for clarity. The green curve in each plot depicts $Q=Q_\mathrm{min}$, the red curves are $Q<Q_\mathrm{min}$, and the black curves are $Q>Q_\mathrm{min}$. The temperatures shown are \textbf{\textsf{a}}, \textbf{\textsf{b}} $T=150$~K, \textbf{\textsf{c}}, \textbf{\textsf{d}} $100$~K, and \textbf{\textsf{e}}, \textbf{\textsf{f}} $75$~K.}
	\label{fig:qmin-low}
\end{figure}

\begin{figure}
	\centering
	\includegraphics[width=\linewidth]{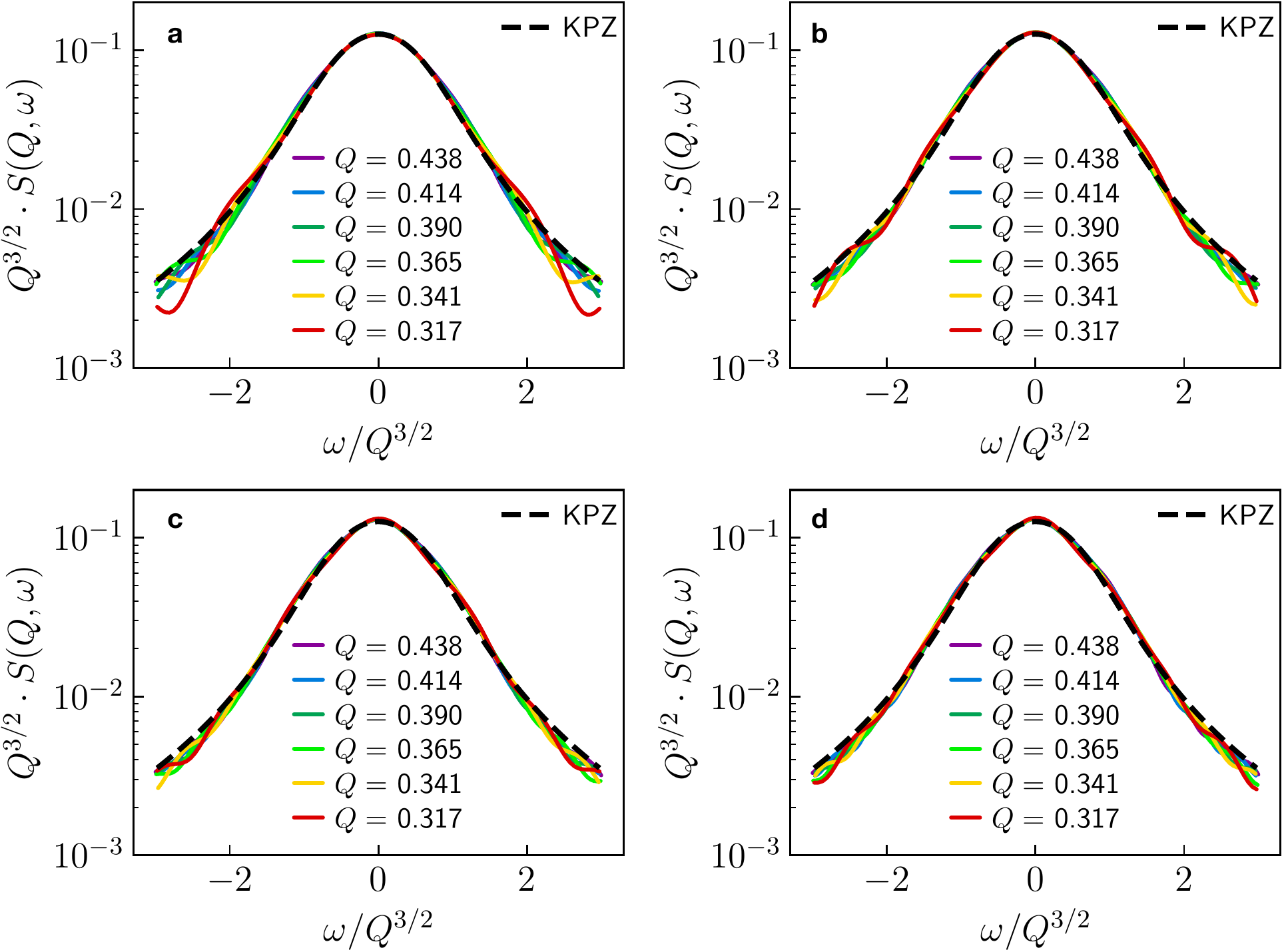}
	\caption{The KPZ scaling function plotted against the numerical data for various values of $Q$ at $T=+\infty$. The different subplots correspond to \textbf{\textsf{a}} $N=1500$, \textbf{\textsf{b}} $N=2000$, \textbf{\textsf{c}} $N=2500$, and \textbf{\textsf{c}} $N=3000$.}
	\label{fig:scalingN}
\end{figure}

In the main text, we include $N=3000$ terms in Eq.~\eqref{eq:Sqw_Cheb} for comparison with experiments. $N/W$ is a pseudo-time parameter in this method, and larger values produce greater $\omega$ resolution. The $Q$ dependence at small $\omega$ is illustrated in Fig.~\ref{fig:qmin-low} and Fig.~\ref{fig:qmin-high}, where we see the simulations are well converged in the region where comparison with experiments is made. Convergence is not found for very small $Q<Q_\mathrm{min}$, which is understood when making the analogy between $N$ and time $t$. Exactly at $Q=0$, then $\hat{S}_{Q=0}^z$ is a conserved quantity, and so for very low $Q$, very large $t$ (and equivalently very large $N$) is needed to accurately resolve the $\omega$ dependence. Nevertheless, the capturing of the KPZ scaling function at $T=+\infty$ is quite robust to $N$, and is illustrated in Fig.~\ref{fig:scalingN}. 

\end{document}